\documentclass[
letterpaper,
english,
reprint,
nofootinbib,
aps,
superscriptaddress,
showpacs,
showkeys,
pra,
floatfix,
]{revtex4-1}

\usepackage{babel,calc,amsmath,amsthm,amssymb,graphicx,subfigure,xcolor,comment}
\usepackage{txfonts}
\usepackage{mathdots}
\usepackage{todonotes}
\usepackage[T1]{fontenc}
\usepackage[unicode=true]{hyperref}
\usepackage{booktabs}
\usepackage{threeparttable}
\usepackage{times}
\usepackage{orcidlink}
\usepackage{CJK}
\hypersetup{
	colorlinks=true,       		
	linkcolor=blue,          	
	citecolor=blue,            
	urlcolor=blue,           	
}

\theoremstyle{definition}

\def\bra#1{\left\langle {#1} \right\rvert}
\def\ket#1{\left\lvert {#1} \right\rangle}

\def\ketOAM#1{{\left\lvert {#1} \right\rangle}^{\mathrm{LG}}}

\begin{document}


\title{Heralded High-Dimensional Photon-Photon Quantum Gate}

\author{Zhi-Feng Liu}
\altaffiliation{These authors contributed equally to this work.}
\author{Zhi-Cheng Ren\orcidlink{0000-0001-5098-8633}}
\altaffiliation{These authors contributed equally to this work.}
\author{Pei Wan}
\author{Wen-Zheng Zhu}
\author{Zi-Mo Cheng}
\author{Jing Wang}
\author{\\Yu-Peng Shi}
\author{Han-Bing Xi}
\affiliation{National Laboratory of Solid State Microstructures, School of Physics, Nanjing University, Nanjing 210093, China}
\affiliation{Collaborative Innovation Center of Advanced Microstructures, Nanjing University, Nanjing 210093, China}

\author{Marcus Huber\orcidlink{0000-0003-1985-4623}}
\affiliation{Atominstitut, Technische Universit{\"a}t Wien, Stadionallee 2, 1020 Vienna, Austria}
\affiliation{Institute for Quantum Optics and Quantum Information - IQOQI Vienna, Austrian Academy of Sciences, Boltzmanngasse 3, 1090 Vienna, Austria}
\author{Nicolai Friis\orcidlink{0000-0003-1950-8640}}
\affiliation{Atominstitut, Technische Universit{\"a}t Wien, Stadionallee 2, 1020 Vienna, Austria}
\affiliation{Institute for Quantum Optics and Quantum Information - IQOQI Vienna, Austrian Academy of Sciences, Boltzmanngasse 3, 1090 Vienna, Austria}

\author{Xiaoqin Gao}
\email[]{xiaoqin.gao.q@gmail.com}
\affiliation{National Research Council of Canada, 1200 Montr{\'e}al Rd, K1A 0R6, Ottawa, ON, Canada}
\affiliation{University of Ottawa, K1N 5N6, Ottawa, ON, Canada}
\author{Xi-Lin Wang\orcidlink{0000-0002-3990-6454}}
\email[]{xilinwang@nju.edu.cn}
\affiliation{National Laboratory of Solid State Microstructures, School of Physics, Nanjing University, Nanjing 210093, China}
\affiliation{Collaborative Innovation Center of Advanced Microstructures, Nanjing University, Nanjing 210093, China}
\affiliation{Hefei National Laboratory, Hefei 230088, China}
\affiliation{Synergetic Innovation Center of Quantum Information and Quantum Physics, University of Science and Technology of China, Hefei 230026, China}
\author{Hui-Tian Wang\orcidlink{0000-0002-2070-3446}}
\email[]{htwang@nju.edu.cn} 
\affiliation{National Laboratory of Solid State Microstructures, School of Physics, Nanjing University, Nanjing 210093, China}
\affiliation{Collaborative Innovation Center of Advanced Microstructures, Nanjing University, Nanjing 210093, China}
\affiliation{Collaborative Innovation Center of Extreme Optics, Shanxi University, Taiyuan 030006, China}


\begin{abstract}
High-dimensional encoding of quantum information holds the potential to greatly increase the computational power of existing devices by enlarging the accessible state space for fixed register size and by reducing the number of required entangling gates. However, qudit-based quantum computation remains far less developed than conventional qubit-based approaches, in particular for photons, which represent natural multi-level information carriers that play a crucial role in the development of quantum networks. 
A major obstacle for realizing quantum gates between two individual photons is the restriction of direct interaction between photons in linear media. 
In particular, essential logic components for quantum operations such as native qudit-qudit entangling gates are still missing for optical quantum information processing. Here we address this challenge by presenting a protocol for realizing an entangling gate---the controlled phase-flip (CPF) gate---for two photonic qudits in arbitrary dimension. We experimentally demonstrate this protocol by realizing a four-dimensional qudit-qudit CPF gate, whose decomposition would require at least 13 two-qubit entangling gates. Our photonic qudits are encoded in orbital angular momentum (OAM) and we have developed a new active high-precision phase-locking technology to construct a high-dimensional OAM beam splitter that increases the stability of the CPF gate, resulting in a process fidelity within a range of $ [0.64 \pm 0.01, 0.82 \pm 0.01]$. Our experiment represents a significant advance for high-dimensional optical quantum information processing and has the potential for wider applications beyond optical system. 
\end{abstract}

\maketitle




While conventional approaches to quantum computing are based on two-dimensional quantum systems---qubits, quantum-information processing with high-dimensional (HD) systems---\textit{qudits}---has increasingly garnered attention~\cite{VertesiPironioBrunner2010, DadaEtAl2011, MalikErhardHuberKrennFicklerZeilinger2016, Martin-Gisin2017, ErhardMalikKrennZeilinger2018, ErhardKrennZeilinger2020, DesignolleSrivastavUolaHerreraValenciaMcCutcheonMalikBrunner2021, HiekkamaekiFickler2021, Chi-Wang2022}. Encoding more quantum information into the same number of information carriers increases their utility for quantum information-processing tasks.
For instance, quantum-communication protocols featuring multilevel systems can benefit from enhanced security against certain types of eavesdropping~\cite{BrussMacchiavello2002}, and key rate generated by quantum key distribution can be improved using HD entanglement~\cite{PivoluskaHuberMalik2018, DodaHuberMurtaPivoluskaPleschVlachou2021}, which have already been distributed via free-space links~\cite{Sit-Karimi2017, BullaHjorthKohoutLangEckerNeumannBittermannKindlerHuberBohmannUrsinPivoluska2023}. 

For quantum computing, HD systems hold the promise to provide increased computational power for existing devices, by making larger state spaces accessible for registers of fixed size~\cite{LanyonEtAl2009, WangHuSandersKais2020, Ringbauer2022, HrmoEtAl2023} and by reducing the number of required costly and error-prone entangling gates~\cite{GaoAppelFriisRingbauerHuber2023}, leading to simplified and more efficient implementations of pertinent algorithms~\cite{KiktenkoNikolaevaXuShlyapnikovFedorov2020, WangHuSandersKais2020, NikolaevaKiktenkoFedorov2021}. 
Despite these advances, the development of qudit-based quantum computation has not yet caught up with conventional qubit-based approaches in many respects. In particular for optical implementations---pivotal for applications in quantum networks, a major challenge lies in the realization of essential logic operations in terms of native qudit-qudit gates~\cite{LanyonEtAl2009, GaoErhardZeilingerKrenn2019}. Together with single-qudit gates, for which photonic implementations are available~\cite{BabazadehEtAl2017}, such entangling gates represent crucial building blocks for scalable optical quantum computing since they are required for universality, i.e., for the construction of arbitrary circuits~\cite{NielsenChuang2010}. Understanding HD two-photon quantum gates is therefore a crucial bottleneck for advancing quantum computation beyond qubits. 

Here, we make a significant step towards overcoming this critical problem: We present a protocol for realizing an entangling two-qudit gate---the \textit{controlled phase-flip} (CPF) gate---between two photonic qudits with arbitrarily high dimensions and we experimentally implement it for two four-dimensional qudits encoded in orbital angular momentum (OAM)~\cite{AllenBeijersbergenSpreeuwWoerdman1992, KrennMalikErhardZeilinger2017}. The CPF gate flips the phase of one of the basis states of a target qudit, conditioned on a specific basis state of the control qudit, and can be used in conjunction with single-qudit gates to realize other two-photon and multi-photon gates~\cite{GaoAppelFriisRingbauerHuber2023}.
 
The main practical obstacle for realizing entangling gates between two or more photons is that photons do not directly interact with each other in linear media. For two-qubit gates between two photons, this restriction has been bypassed either by destructive measurements and post-selection~\cite{OBrienPrydeWhiteRalphBranning2003, PittmanJacobsFranson2002, LangfordWeinholdPrevedelReschGilchristOBrienPrydeWhite2005, OkamotoHofmannTakeuchiSasaki2005}, requiring an output measurement for gate-success verification and thus limiting further experimentation, or by heralding~\cite{GasparoniPanWaltherRudolphZeilinger2004, ZhaoZhangChenZhangDuYangPan2005, HuangRenZhangDuanGuo2004, ZeunerSharmaTillmannHeilmannGraefeMoqanakiSzameitWalther2018, BaoChenZhangYangZhangYangPan2007, LiGuQinWuYouWangSchneiderHoeflingHuoLuLiuLiPan2021}, offering non-destructive gate-operation feedback without an output measurement. Non-destructive gates obtained via heralding can be combined into gate operations on larger quantum systems, enabling efficient linear quantum computation. However, this approach usually requires auxiliary single photons~\cite{BaoChenZhangYangZhangYangPan2007, LiGuQinWuYouWangSchneiderHoeflingHuoLuLiuLiPan2021} or entangled photon pairs~\cite{GasparoniPanWaltherRudolphZeilinger2004, ZhaoZhangChenZhangDuYangPan2005, HuangRenZhangDuanGuo2004, ZeunerSharmaTillmannHeilmannGraefeMoqanakiSzameitWalther2018, GaoAppelFriisRingbauerHuber2023}. Precisely preparing the latter is challenging, which negatively impacts gate fidelities and hinders high-precision quantum computation. 
As a result, the implementation of non-destructive HD CPF gates using single photons---demonstrated here for the first time---is essential for advancing HD quantum-information processing. 

Our implementation using OAM profits from the stability and reliability provided by this degree of freedom (DoF) for preparing entangled states~\cite{MairVaziriWeihsZeilinger2001, WangEtAl2018}. In addition, we introduce a novel phase-locking scheme to the multi-path interference loops, which is very stable in our experiment and shows promise for wider applications in manipulating spatially modulated HD quantum states.


\section*{Theory of The CPF Gate}

Considering a pair of $d$-dimensional quantum systems, we label them as $1$ and $4$, for reasons that will become clear shortly, associated with Hilbert spaces $\mathcal{H}_{1}$ and $\mathcal{H}_{4}$, respectively. With respect to a pair of selected orthonormal bases $\bigl\{\ket{m}_{1}\bigr\}$ $(m=0,...,d-1)$ and $\bigl\{\ket{n}_{4}\bigr\}$ $(n=0,...,d-1)$, a general pure initial state (which can be a product state or an entangled state) takes the form of $\ket{\psi}_{14} = \sum_{m,n=0}^{d-1}c_{m,n} \ket{m}_{1} \ket{n}_{4}$.
The action of the unitary CPF gate $U_{\mathrm{CPF}}$ is to apply a phase shift of~$\pi$ if both systems are in the state $\ket{d - 1}$, and no phase shift otherwise, 
\begin{align}
    U_{\mathrm{CPF}} = \sum \limits_{m,n=0}^{d-1} \ket{m,n} \bra{m,n} - 2 \ket{d - 1, d - 1} \bra{d - 1,d - 1},
    \label{eq:CPF gate}
\end{align}
where $\ket{m,n} = \ket{m} \otimes \ket{n} \in \mathcal{H}_{1} \otimes \mathcal{H}_{4}$. Like its two-dimensional counterpart, the controlled-$Z$ gate, the CPF gate is symmetric with respect to the exchange of the two systems, which are usually designated as control and target, here, systems 1 and 4, respectively.

\begin{figure*}[t]
\includegraphics [width= 0.90\textwidth]{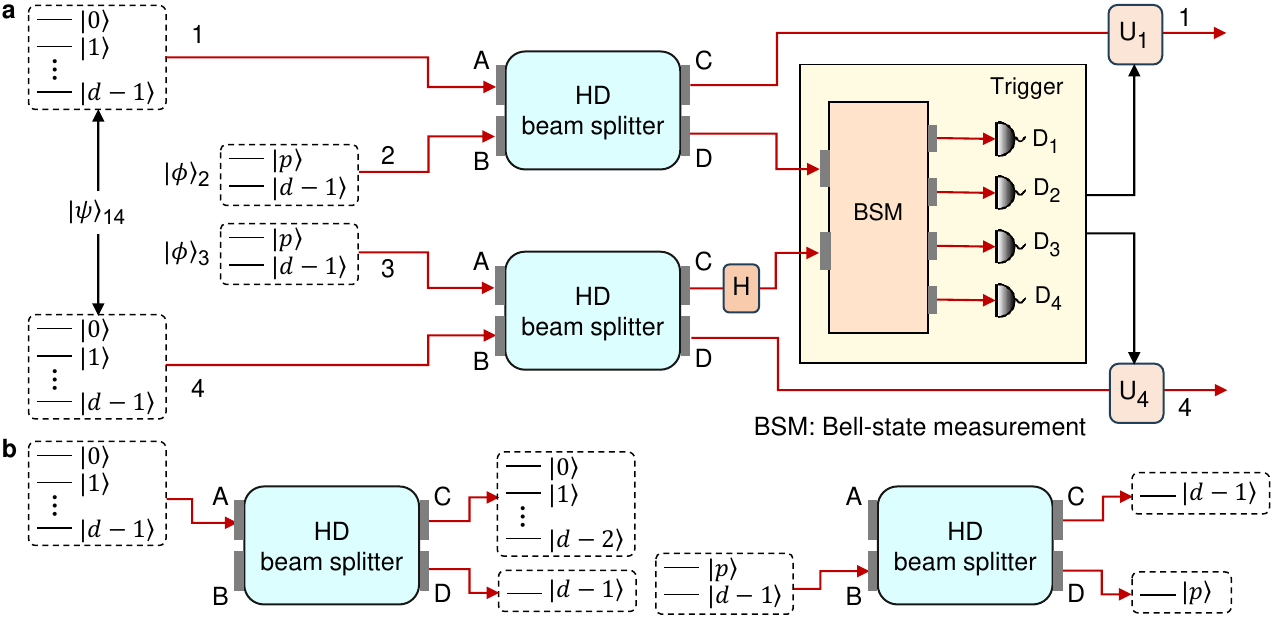}
\centering
	\caption{ \textbf{Protocol for realizing two-qudit CPF gate}.
		{\textbf{a}} Input photons~$1$ and~$4$ encoding two qudits in the initial joint state $\ket{\psi}_{14}$ are respectively paired on two HD beam splitters with two auxiliary photons, labelled~$2$ and~$3$ in the same state $\ket{\phi}_{2} = \ket{\phi}_{3} = \frac{1}{\sqrt{2}} (\ket{p} + \ket{d - 1})$ in two-dimensional subspaces spanned by $\ket{p}$ and $\ket{d - 1}$ for $p$ satisfying $0 \leq p < d-1$. \textbf{b} The HD beam splitter with two input ports (A and B) and two output ports (C and D) has the following operational functions: photons in the state $\ket{d - 1}$ entering from port~A (B) exit from port~D (C), i.e. $\rm{A}\!\rightarrow\!\rm{D}$ and $\rm{B}\!\rightarrow\!\rm{C}$ for photons in $\ket{d - 1}$; while photons in all other states entering from port~A (B) exit from port~C (D), i.e. $\rm{A}\!\rightarrow\!\rm{C}$ and $\rm{B}\!\rightarrow\!\rm{D}$ for photons not in $\ket{d - 1}$.
		After applying a Hadamard gate (H) acting on the two-dimensional subspace spanned by the states $\ket{p}$ and $\ket{d - 1}$ of the photons exiting port~D of the second HD beam splitter, a joint Bell-state measurement (BSM) is carried out on the photons exiting from the two ports labelled~D, and suitable local unitary corrections $U_{1}$ and $U_{4}$ are applied to the photons in the two ports labelled~C, depending on the outcome of the BSM. We finally post-select the events that one and only one photon has been detected in both ports~C and~D. See the Appendix for more details.}
	\label{fig1}
\end{figure*}

To realize an HD CPF gate, we introduce two auxiliary systems labelled as 2 and 3, respectively, both are initialized in the same state of $\ket{\phi}_{2}=\ket{\phi}_{3}=\frac{1}{\sqrt{2}}(\ket{p}+\ket{d - 1})$ in two-dimensional subspaces spanned by $\ket{p}$ and $\ket{d-1}$ for $p$ satisfying $0 \leq p < d-1$. As shown in Fig.~\ref{fig1}, input photons~$1$ and~$4$ are paired with two auxiliary photons~$2$ and~$3$ on two HD beam splitters, respectively. As shown in Fig.~\ref{fig1}, when photons in the state $\ket{d - 1}$ enter from port~A (B) of the HD beam splitter, they will exit from port~D (C); while photons in the other states $\ket{0},\ldots,\ket{d - 2}$ entering from port~A (B), will exit from port~C (D). We post-select on one and only one photon exiting from each output port, so that the photons can again be labelled $1$, $2$, $3$, and $4$, according to their output ports, see Fig.~\ref{fig1}.
After applying a Hadamard gate (H) in the subspaces spanned by $\ket{p}_{3}$ and $\ket{d - 1}_{3}$ of photon~$3$, we obtain the four-photon state
\begin{align}
  \left| \psi \right\rangle_{\mathrm{f}} & =    U_{\mathrm{CPF}}\ket{\psi}_{14} \ket{\Phi^{+}}_{23}
  + {U_{1}}U_{\mathrm{CPF}}\ket{\psi}_{14} \ket{\Phi^{-}}_{23} \nonumber \\
  & \ \ \ \ + {U_{4}}U_{\mathrm{CPF}}\ket{\psi}_{14} \ket{\Psi^{+}}_{23} + {U_{1}}{U_{4}}U_{\mathrm{CPF}}\ket{\psi}_{14} \ket{\Psi^{-}}_{23}.
   \label{eq:four-photon state}
\end{align} 
To obtain the output state $U_{\mathrm{CPF}}\ket{\psi}_{14}$ for photons 1 and 4 in Eq.~(\ref{eq:four-photon state}), we first need to jointly measure photons~$2$ and~$3$ with respect to the Bell basis consisting of the states
\begin{subequations}
\begin{align}
    \ket{\Phi^{\pm}}_{23} & = \frac{1}{\sqrt{2}}\bigl( \ket{p}_{2} \ket{p}_{3} \pm \ket{d - 1}_{2} \ket{d - 1}_{3} \bigr), \\
    \ket{\Psi^{\pm}}_{23} & = \frac{1}{\sqrt{2}}\bigl( \ket{p}_{2} \ket{d - 1}_{3} \pm \ket{d - 1}_{2} \ket{p}_{3} \bigr),
\end{align}
\end{subequations}
and then apply suitable local unitary corrections ($U_{1}= I - 2\ket{d - 1} \bra{d - 1}_{1}$, $U_{4}= I - 2\ket{d - 1} \bra{d - 1}_{4}$, or $U_{1}U_{4}$, depending on the outcome of the Bell-state measurement, where $I$ is an identity matrix). Finally, the two-photon output state (for photons~$1$ and~$4$) is $U_{\mathrm{CPF}} \ket{\psi}_{14}$, where $U_{\mathrm{CPF}}$ is the desired CPF gate from Eq.~(\ref{eq:CPF gate}). A detailed derivation is given in Sec.~I of the Appendix.

Finally, we note that the fact that the auxiliary modes are initialized in uniform superpositions of two states, which exit from two different ports of the HD beam splitter, respectively, means that the probability for obtaining one and only one photon in each output mode of the HD beam splitter is $1/2$, independently of the input state of photons~$1$ and $4$. The probability of successfully realizing two such HD multiports is thus $1/4$. In addition, only two of the four Bell states can be distinguished unambiguously in our experiment, leading to an overall theoretical process efficiency of the CPF gate of $1/8$. However, such a process efficiency of $1/8$ is independent of $d$, and hence does not decrease for higher dimensions.


\section*{Experiment}

We now describe our experimental realization of the CPF gate for two four-dimensional ($d=4$) qudits encoded in \textit{orbital angular momentum} (OAM) of single photons~\cite{AllenBeijersbergenSpreeuwWoerdman1992, KrennMalikErhardZeilinger2017}. 
We represent the four computational-basis states of $\ket{0}$, $\ket{1}$, $\ket{2}$, and $\ket{3}$ by four OAM states $\ketOAM{-2}$, $\ketOAM{-1}$, $\ketOAM{0}$, and $\ketOAM{+1}$ (i.e. $\ketOAM{-2}\!\rightarrow\!\ket{0}$, $\ketOAM{-1}\!\rightarrow\!\ket{1}$, $\ketOAM{0}\!\rightarrow\!\ket{2}$, $\ketOAM{+1}\!\rightarrow\!\ket{3}$), respectively, where $\ketOAM{\ell}$ denotes the photon being in a Laguerre-Gauss mode with radial index zero and azimuthal quantum number~$\ell$. In addition, we assume that all photons are initially prepared with horizontal polarization.

\begin{figure}[!ht]
\centering
\includegraphics[width=0.96\linewidth]{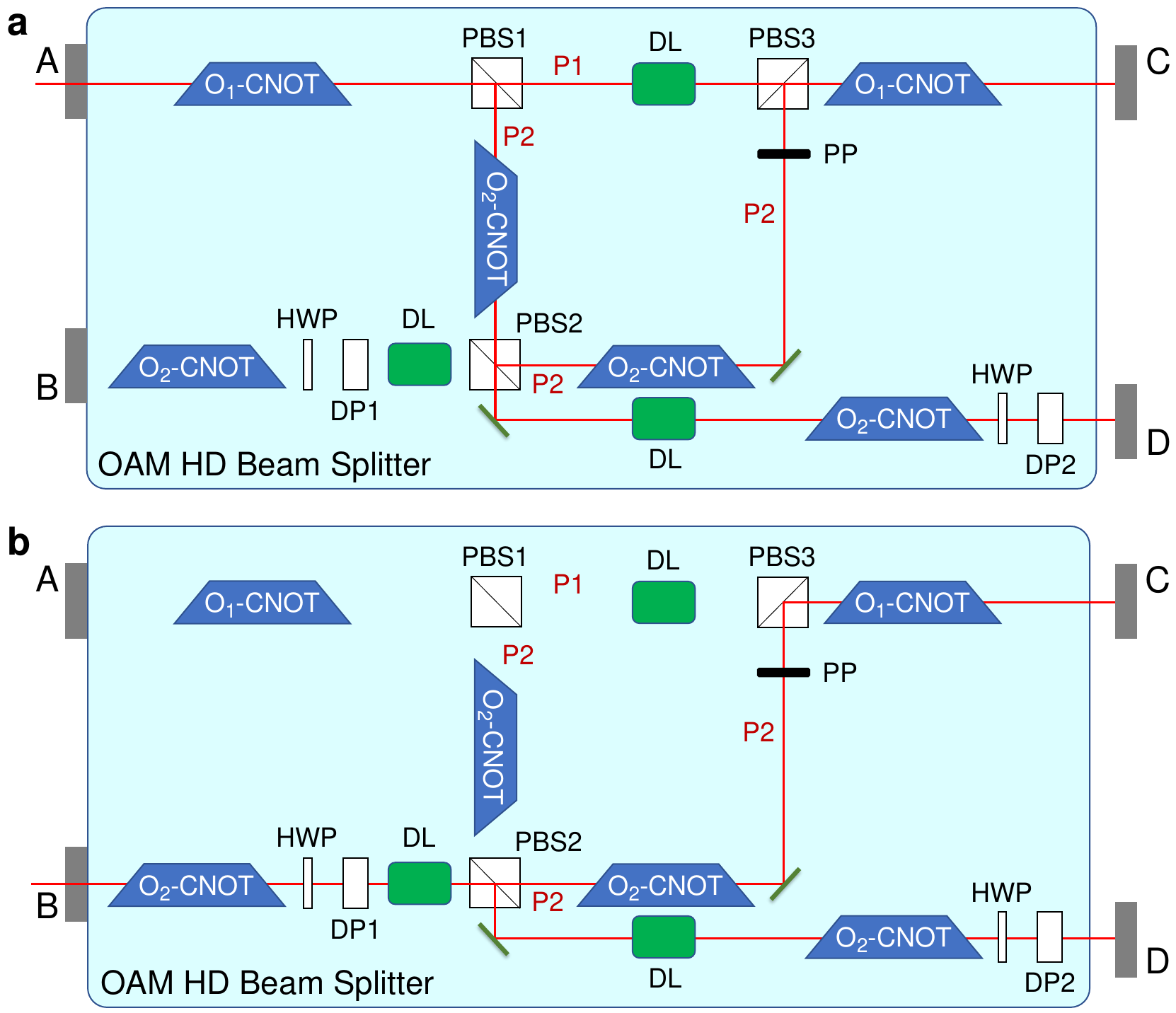}
	\caption{Structural diagram and operation principle of OAM HD beam splitter with two input ports (A and B) and two output ports (C and D) for $d=4$. \textbf{a} Situation when photons enter from port A; photons in the OAM state $\ketOAM{+1}$ exit from port D, while photons in the other three states ($\ketOAM{-2} $, $\ketOAM{-1}$, $\ketOAM{0}$) exit from port C. \textbf{b} Situation when photons enter from port B; photons in the OAM state $\ketOAM{+1}$ exit from port C, while photons in the other three states ($\ketOAM{-2} $, $\ketOAM{-1}$, $\ketOAM{0}$) exit from output port D. To achieve this, O$_{k}$-CNOTs (for $k=1,2$) correlate polarization and OAM. Step-by-step state transformations are provided in Sec.~IIC of the Appendix. PBS, polarizing beam splitter; HWP, half-wave plate; DP, Dove prism; PP, phase plate; DL, delay line. P1 and P2 label optical paths.}
	\label{Fig2}
\end{figure}

The centrepiece of the CPF-gate implementation using OAM is the OAM beam splitter~\cite{MalikErhardHuberKrennFicklerZeilinger2016, ZhangQiZhouChen2014} that separates OAM modes into different paths; as described above, photons in the mode $\ketOAM{+1}$ are directed onto one path, while photons in all other modes ($\ketOAM{-2}$, $\ketOAM{-1}$, $\ketOAM{0}$) are directed to another.  
An arrangement of linear-optical elements realizing such an operation in the OAM DoF is shown in Fig.~\ref{Fig2}.  
We employ OAM-based O$_{k}$-CNOT gates~\cite{WangEtAl2018} with different orders $k=1,2$ in combination with polarizing beam splitters (PBSs), half-wave plates (HWPs), quarter-wave plates (QWPs), Dove prisms (DPs), phase plates (PPs), delay lines (DLs), and mirrors. 

\begin{figure*}[ht!]
\includegraphics [width= 0.93\textwidth]{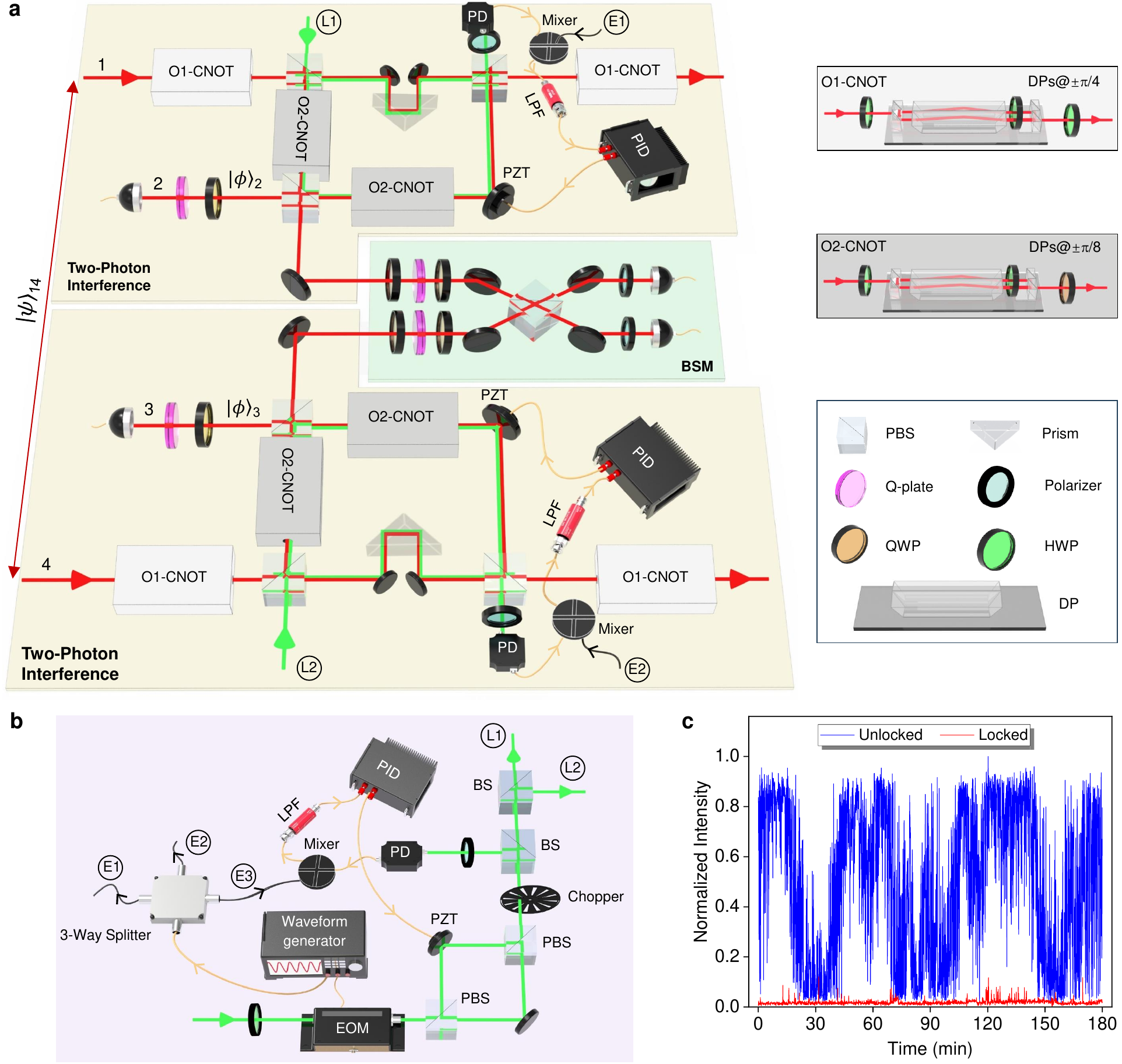}
\centering
\caption{\textbf{Experimental setup for heralded four-dimensional OAM CPF gate.} 
\textbf{a} Experimental setup. The central elements of the setup are two stages of two-photon interference (light yellow background) and the Bell-state measurement (BSM, light green background). In the two-photon interference, O$_{k}$-CNOTs for $k=1,2$ denote single-photon entangling gates that act on joint polarization-OAM states, and realize HD OAM beam splitters in conjunction with polarizing beam splitters (PBSs), as explained in more detail in Sec.~IIC of the Appendix. Following the two-photon interference a Bell-state measurement (BSM) is realized to herald the OAM CPF gate. 
To stabilize the setup, we introduced a new phase-locking method by adding two modulated locking-laser beams to the control and target paths, respectively. The piezoelectric ceramics (PZTs) compensate for the phase change. 
\textbf{b} Preparation of phase-locking laser. An electro-optic modulator (EOM) is used to modulate the commercial phase-locking laser. The MZ interferometer is used to compensate for the slow temperature change that affects the stability of the EOM. A chopper ensures coincidence counting without influence from the locking laser. \textbf{c} Locking result over three hours. We use the input state $\frac{1}{\sqrt 2} \left( \ket{H} + \ket{V} \right) \otimes \ketOAM{+1}$ for both the control and target qudits encoded in photons~$1$ and~$4$, respectively, and we measure the corresponding outputs in the basis $\frac{1}{\sqrt 2} \left( \ket{H} - \ket{V} \right) \otimes \ketOAM{+1}$. Then we compare the output with $\frac{1}{\sqrt 2} \left( \ket{H}  + \ket{V} \right) \otimes \ketOAM{+1}$.
The setup is much more stable after locking the phase (red line) compared with the unlocked case (blue line). Acronyms: QWP, quarter-wave plate; HWP, half-wave plate; PBS, polarizing beam splitter; BS, beam splitter; DP, Dove prism; PID, proportional-integral-derivative phase compensation for improving the performance of the servo control system; PD, photodetector; LPF, low-pass filter.}
	\label{fig3}
\end{figure*}

Each such O$_{k}$-CNOT gate consists of an interferometer (which itself includes two DPs and a HWP) between two HWPs (or a HWP and a QWP). The O$_{k}$-CNOTs with different orders~$k$ have different orientation angles of the DPs with respect to the HWPs, as illustrated in the insets of Fig.~\ref{fig3}\textbf{a}. The O$_{k}$-CNOT gate acts as an entangling gate on the joint polarization-OAM states of single photons, switching the polarization of the respective photons conditioned on their OAM states. For photons in the OAM state $\ketOAM{\ell}$ and horizontal ($\ket{H}$) or vertical ($\ket{V}$) polarization, the O$_{1}$-CNOT acts as
\begin{subequations}
\begin{align}
\text{O$_{1}$-CNOT} \ket{H} \ketOAM{\ell} & = \phantom{-}e^{-i \ell \pi/2}
\begin{cases}
\ket{H}\ketOAM{-\ell} & \! \text{even $\ell$},\\
\ket{V}\ketOAM{-\ell} & \! \text{odd $\ell$},
\end{cases} \\
\text{O$_{1}$-CNOT} \ket{V} \ketOAM{\ell} & = - e^{-i \ell \pi/2}
\begin{cases}
\ket{V}\ketOAM{-\ell} & \! \text{even $\ell$},\\
\ket{H}\ketOAM{-\ell} & \! \text{odd $\ell$}.
\end{cases}
\end{align}
\label{eq:4}
\end{subequations}
While O$_{2}$-CNOT leads to $\ell$-dependent superpositions of $\ket{H}$ and $\ket{V}$ in addition to changing the sign of $\ell$. 
In our setup we use two types of O$_{k}$-CNOTs in conjunction with PBSs to sort photons with different $\ell$ into different paths.
For further details on the composition of the O$_{k}$-CNOTs, OAM HD beam splitters, and derivations of their transformations, we refer to Sec.~IIC of the Appendix. 

In our experiment, the four input photons originate from two pairs, photons~$1$ and~$2$ as well as photons~$3$ and~$4$, produced by a type-II spontaneous parametric down-conversion (SPDC) process~\cite{WangEtAl2016, Zhong-Pan2018}. A femtosecond pulsed laser with a repetition rate of 80 MHz and a central wavelength of 390 nm successively pumps two single $2$-mm-thick $\beta$-barium borate (BBO) crystals.
Signal and idler photons are collected by single-mode fibers (SMFs) and transported to the input ports of the photonic CPF gate. During transmission through the SMF, all the non-Gaussian modes generated in the SPDC process are filtered out, and the initial polarization of all photons is set to $\ket{H}$ using a PBS.

Photons~$1$ and~$4$ encode the control and target qudits, respectively. Both are prepared in several different input states to test the implementation of the CPF gate.
These states include: four computational-basis states $\ketOAM{-2}$, $\ketOAM{-1}$, $\ketOAM{0}$, and $\ketOAM{+1}$, referred to as the $Z$ basis; four equally weighted superpositions of pairs of OAM states with azimuthal quantum numbers of the same parity, $\frac{1}{{\sqrt 2 }} \bigl(\ketOAM{-2}\pm\ketOAM{0}\bigr)$ and $\frac{1}{{\sqrt 2 }} \bigl(\ketOAM{-1} \pm \ketOAM{+1}\bigr)$, which we refer to as the $X$ basis; as well as two additional equally weighted superpositions of computational-basis states, $\frac{1}{{\sqrt 2 }} \bigl(\ketOAM{-1}+\ketOAM{0}\bigr)$ and $\frac{1}{{\sqrt 2 }} \bigl(\ketOAM{0}+\ketOAM{+1}\bigr)$.  
These states are generated from the initial Gaussian mode ($\ell=0$) with horizontal polarization by selecting suitable combinations of optical components, including HWPs, QWPs, q-plates (QPs)~\cite{MarrucciManzoPaparo2006}, and spiral phase plates (SPPs)~\cite{KotlyarAlmazovKhoninaSoiferElfstromTurunen2005} (see Sec.~IIA of the Appendix for more details).

For the state preparation of auxiliary photons 2 and 3, we note that the input states $\ket{\phi}_{2} =\ket{\phi}_{3}= \frac{1}{\sqrt{2}} \ket{H} (\ket{p}+\ket{d-1}) = \frac{1}{\sqrt{2}} \ket{H} (\ketOAM{-1} + \ketOAM{+1})$ with $p=1$ at the input port B of the OAM HD beam splitters are there immediately transformed into $\ket{\phi'}=\frac{1}{{\sqrt 2 }} \left( \ket{V} \ketOAM{-1} + \ket{H} \ketOAM{+1} \right)$ by sequences of O$_{2}$-CNOTs, HWPs, DPs, and DLs.
In our experiment, a schematic of which is shown in Fig.~\ref{fig3}\textbf{a}, we have therefore simplified the setup by directly preparing $\ket{\phi'}$ using a QP and a QWP (see Sec.~IIB of the Appendix for details).

Photon~$1$ and the auxiliary single photon~$2$, as well as the second auxiliary photon~$3$ and photon~$4$ then pairwise interfere at two OAM HD beam splitters, respectively. 
These two instances of two-photon interference in Mach-Zender (MZ) interferometers with two single input photons form the core building blocks of our setup.
The key to the successful operation of our CPF gate is the phase stability of the MZ interferometers.
To this end, we have developed an active phase-locking scheme for OAM, as shown in Fig.~\ref{fig3}\textbf{b}, which allows us to lock the MZ interferometer to an arbitrary phase using a phase-locking laser, thus stabilizing the setup. 
As explained in the detailed description in Sec.~III of the Appendix, we use an electro-optic modulator (EOM)~\cite{WuHuangYangLiuChen2019} to modulate the locking laser and introduce a set of MZ interferometers to compensate for the effect of slow temperature change. 
Such a control results in a relatively stable phase-locking laser with a final output of $\ket{H}$ and $\ket{V}$.
We introduced the locking laser through the PBS into the two MZ interferometers. Then, the two polarization components of the locking laser enter the two paths of the interferometers separately. The DC signal received by the photodetector controls the switching of the single-photon avalanche detectors, ensuring that the stability of the two MZ interferometers is maintained over three hours by the locking laser, as shown in Fig.~\ref{fig3}\textbf{c}. A chopper is inserted into the optical path before the locking laser enters the two MZ interferometers, to prevent the locking laser from disturbing the coincidence counting of photons. 

The photons exiting the MZ interferometer via the paths corresponding to output ports D in Fig.~\ref{Fig2} transfer their OAM information to the polarization DoF and then enter a standard BSM, with one of the photons first being subjected to a Hadamard gate. As discussed in more detail in Sec.~IID of the Appendix, the experimental setup for performing these operations has been simplified with respect to Fig.~\ref{Fig2} by combining operations in ports D of the OAM HD beam splitters, Hadamard gate, and the BSM. 

We measure the OAM information of photon~$1$ 
by first transferring the information from the OAM DoF to the polarization DoF~\cite{WangEtAl2018}, then making a projection measurement using a polarization analyzer.
The projection measurement of photon~$4$ is executed using a spatial light modulator (SLM)~\cite{AgnewLeachMcLarenRouxBoyd2011}. The two photons, along with the auxiliary single photons that undergo the BSM projection to $\ket{\Phi^{+}}_{23}$, are collected by SMFs and directed to the avalanche photodiode detectors (APD) for four-fold coincidence counting.

\begin{figure}[ht!]
\centering
\includegraphics[width=0.9\linewidth]{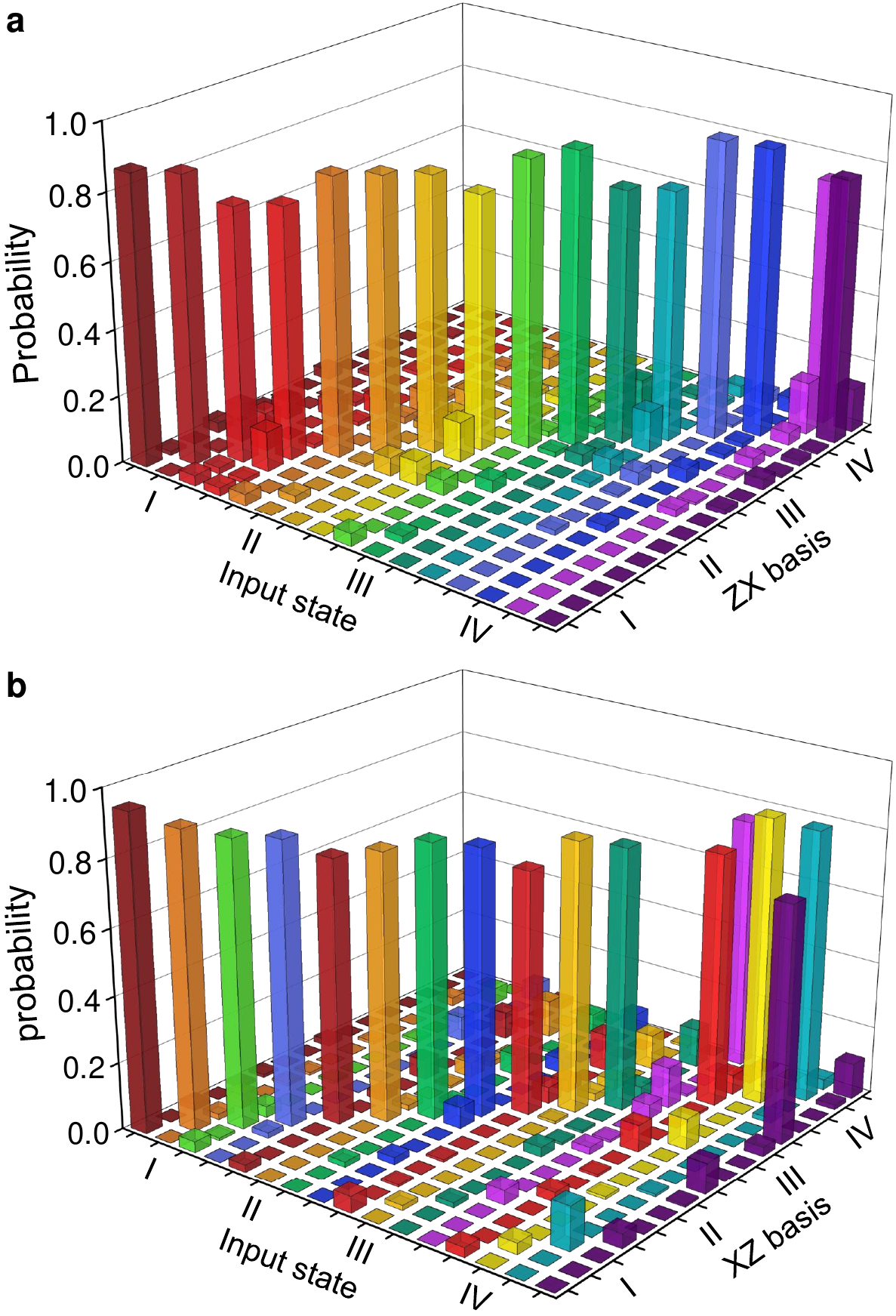}
\caption{\textbf{Experimental results to calculate the process fidelity of the realized heralded CPF gate.} We prepared two sets of two-photon product states for photons~$1$ and~$4$ in the joint $ZX$ and $XZ$ bases in four-dimensional OAM space, respectively, i.e., one photon is prepared in the $Z$ basis and the other in the $X$ basis. We then measured the four-photon coincidences of two-photon states after applying the CPF gate using the same measurement bases ($ZX$ and $XZ$) as for the input states.	
The coloured bars show the probabilities estimated from the measured four-photon coincidences of the two-photon state $U_{\mathrm{CPF}} \ket{\psi}_{14}$ after the CPF gate. \textbf{a} shows the case where the two-photon input state is in the $ZX$ basis and the measurement basis is the same $ZX$ basis. \textbf{b} shows the case where the two-photon input state is in the $XZ$ basis and the measurement is carried out with respect to the same $XZ$ basis. As shown in \textbf{a}, out of the 16 states of the $ZX$ basis, only the states $\ketOAM{+1}_1\otimes \frac{1}{\sqrt 2} \left( {\ketOAM{-1}_4 \pm \ketOAM{+1}_4} \right) $ are flipped to $\ketOAM{+1}_1 \otimes \frac{1}{\sqrt 2} \left( {\ketOAM{-1}_4 \mp \ketOAM{+1}_4} \right) $, respectively, while the other states are unchanged. As shown in \textbf{b}, out of the 16 states in the $XZ$ basis, only the states $ \frac{1}{\sqrt 2} \left( {\ketOAM{-1}_1 \pm \ketOAM{+1}_1} \right) \otimes \ketOAM{+1}_4$ are flipped to $ \frac{1}{\sqrt 2} \left( {\ketOAM{-1}_1 \mp \ketOAM{+1}_1} \right)\otimes\ketOAM{+1}_4$, respectively, while the other states are unchanged.
See Figs. S1 and S2 in the Appendix for the detailed state representation for each element of \textbf{a} and \textbf{b}.}
	\label{fig4}
\end{figure}

\section*{Results}

To evaluate the performance of the CPF gate, we used an efficient method~\cite{Hofmann2005} to estimate upper and lower bounds on the process fidelity $F$. These bounds can be obtained by selecting two complementary sets of bases as inputs to the CPF gate. Specifically, we prepared two sets of two-photon product states for photons~$1$ and~$4$ in the joint $ZX$ and $XZ$ bases in four-dimensional OAM space, respectively, i.e., one photon is prepared in the $Z$ basis and the other in the $X$ basis. We then measured the four-photon coincidences of two-photon states resulting from the CPF-gate operation using the same measurement bases ($ZX$ and $XZ$) as the input states (see Sec.~IV of the Appendix for more details). 
Accordingly, Figs.~\ref{fig4}\textbf{a}~and~\ref{fig4}\textbf{b} show the probabilities estimated from the measured four-photon coincidences for the two sets of two-photon input states (in the $ZX$ basis and in the $XZ$ basis), respectively.
From the results shown in Figs.~\ref{fig4}, we can determine the average fidelity $F_{ZX}$ ($ F_{XZ}$) by calculating the average probability of the expected outputs over all the 16 possible inputs, and we obtain the measured fidelities of $F_{ZX}=0.82 \pm 0.01$ and $F_{XZ} = 0.82 \pm 0.01$. Since the upper and lower bounds of $F$ are defined as $F \in [F_{ZX}+F_{XZ}-1, \min \left\{ {F_{ZX},F_{XZ}} \right\}]$, we achieve a process fidelity of $F \in [0.64 \pm 0.01, 0.82 \pm 0.01]$ for our CPF gate. The process fidelity $F$ is closely linked to the entanglement capacity of the CPF gate, signifying its ability to generate entangled states from separable product states~\cite{Hofmann2005}. Our process fidelity results notably surpass the threshold of 0.5, confirming the capability of our CPF gate to generate entanglement.

\begin{figure}[!htp]
	\centering
	\includegraphics[width=\linewidth]{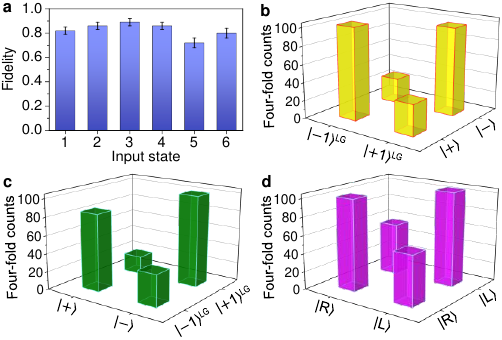}
		\caption{\textbf{Operation of the Heralded CPF gate on two photons with seven input OAM superposition states.} The seven OAM-superposition input states are listed in Sec.~IV of the Appendix. The output of the first six states are product states and the output of the seventh state is an entangled state. \textbf{a} shows the experimentally obtained fidelity of the first six product states. \textbf{b}-\textbf{d} show the measured four-fold counts in various bases to calculate the fidelity of the entangled state for the seventh input state. The seventh input product state is  $\ket{+}_{1}\ket{+}_{4}$ and the corresponding output state is the entangled state $\frac{1}{ 2 }(\ketOAM{-1}_{1}\ketOAM{-1}_{4} + \ketOAM{-1}_{1}\ketOAM{+1}_{4} + \ketOAM{+1}_{1}\ketOAM{-1}_{4} -  \ketOAM{+1}_{1}\ketOAM{+1}_{4})$. To calculate its fidelity, we implement a measurement of two photons in the eigenbasis of $\sigma_{z} \otimes \sigma_{x}$ in \textbf{b}, $\sigma_{x} \otimes \sigma_{z}$ in \textbf{c}, and $\sigma_{y} \otimes \sigma_{y}$ in \textbf{d}, where $\ket{\pm}=\frac{1}{\sqrt{2}}\bigl(\ketOAM{-1}\pm\ketOAM{+1}\bigr)$, $\ket{R}= \tfrac{1}{\sqrt 2 }\bigl(\ketOAM{-1} +i \ketOAM{+1}\bigr)$, $\ket{L}= \frac{1}{\sqrt 2 }\bigl( \ketOAM{-1} - i \ketOAM{+1}\bigr)$.}
	\label{fig5}
\end{figure}

We further tested the CPF gate's ability to produce the desired output for various OAM superposition states. The seven input superposition states we selected are listed in the Sec.~IV of the Appendix. For the first six OAM superposition states, the output states remain unchanged with respect to the product state, with the fidelity shown in Fig.~\ref{fig5}\textbf{a}. When we input the seventh superposition state $\tfrac{1}{ \sqrt{2} }\bigl(\ketOAM{-1}_{1} + \ketOAM{+1}_{1}\bigr) \otimes \tfrac{1}{ \sqrt{2} }\bigl(\ketOAM{-1}_{4} + \ketOAM{+1}_{4}\bigr)$, the CPF gate generates an entangled state $\tfrac{1}{ 2 }\bigl(\ketOAM{-1}_{1}\ketOAM{-1}_{4} + \ketOAM{-1}_{1}\ketOAM{+1}_{4} + \ketOAM{+1}_{1}\ketOAM{-1}_{4} -  \ketOAM{+1}\ketOAM{+1}_{4} \bigr)$. The transition from a two-photon product state to an entangled state is a distinctive feature of the CPF gate. To measure the fidelity of the output entangled state, we measure the expectation values of $\sigma_{z} \otimes \sigma_{x}$, $\sigma_{x} \otimes \sigma_{z}$ and $\sigma_{y} \otimes \sigma_{y}$, where $\sigma_{x,y,z}$ are three usual Pauli matrices in the two-dimensional space spanned by $\ketOAM{-1}$ and $\ketOAM{+1}$, as shown in Fig.~\ref{fig5}\textbf{b}-\textbf{d}, respectively. The fidelity obtained for the entangled state is $0.59 \pm 0.02$.

\section*{Conclusion}

We have presented a strategy for implementing a heralded CPF gate, and  demonstrated experimentally the first successful realization in local dimensions four for the OAM degrees of freedom, leveraging a novel phase-locking technique. Our experimental results follow recent theoretical proposals for realizing CPF gates using the interaction of photons with ions trapped in a cavity~\cite{ChenTangCaiNoriXia2024}, and demonstrate the notable reliability of the CPF gate in producing both product and entangled states. Our flexible phase-locking technology developed for our experiment can find a variety of new applications in areas such as linear optical quantum information processing and quantum communication~\cite{Kimble2008, VitelliSpagnoloAparoSciarrinoSantamatoMarrucci2013, GaoKrennKyselaZeilinger2019}.

The outcomes of our study are poised to contribute to a myriad of high-dimensional quantum information tasks, such as the analysis of Bell states~\cite{BarreiroWeiKwiat2008, Wang-Pan2015, WilliamsSadlierHumble2017} and high-dimensional quantum teleportation~\cite{Luo-ZeilingerPan2019, Reimer-Morandotti2019, Hu-Guo2020}.
The CPF gates, serving as the foundation for numerous entangling gates, can be applied to diverse high-dimensional quantum state transformations. When combined with single-photon high-dimensional gates, this provides a way for accomplishing a range of complex high-dimensional quantum computational tasks, such as quantum error correction~\cite{Terhal2015, BocharovRoettelerSvore2017, MiaoEtAl2023}. Our results thus mark a significant advance for high-dimensional quantum information processing.






\










\begin{center}
\textbf{APPENDIX}	
\end{center}

\
	
\setcounter{equation}{0} 
\renewcommand{\theequation}{A\arabic{equation}}
	
\setcounter{figure}{0} 
\renewcommand{\thefigure}{A\arabic{figure}}
	
In this Appendix we present additional details and explanations for the protocol to realize a controlled phase-flip (CPF) gate and its implementation. In Sec.~\ref{sec:CPF in any dimension} we give a step-by-step derivation of the protocol for realizing a CPF gate in arbitrary dimensions is presented. In Sec.~\ref{sec:OAM implementation} we then discuss the specific implementation of the protocol in local dimension $d=4$ using orbital angular momentum (OAM) of photons. In Sec.~\ref{sec:arbitrary quantum interference phase locking} we provide more information on the phase-locking technique we developed to stabilize the lasers, while in Sec.~\ref{sec:bases used to estimate the fidelity of the CPF gate} we provide a detailed list of the measurements carried out to estimate the fidelity of the CPF gate in our experiment.
	
	
\section{Protocol Implementing a CPF Gate in Arbitrary Dimensions}\label{sec:CPF in any dimension}
	
In this section we present a detailed description of the protocol for implementing the two-photon CPF gate for arbitrary local dimensions $d$. 
To this end, we consider four photons, of which photons~$1$ and~$4$ encode the control and target qudits, respectively, such that a general pure input state for the CPF gate is represented by a joint state of the form of
	$\ket{\psi}_{14}=\sum_{m,n =0}^{d-1}c_{m,n }\ket{m}_{1} \ket{n }_{4}$, where the amplitudes satisfy ${{\sum} _{m,n =0}^{d-1}\left| {{c_{m,n }}} \right|^2} = 1$. 
In addition, we also need to consider two auxiliary photons, labelled~$2$ and~$3$, respectively, which are prepared in the same states of $\ket{\phi}_{2}=\ket{\phi}_{3}=\frac{1}{\sqrt{2}}(\ket{p}+\ket{d - 1})$ in two-dimensional subspaces spanned by $\ket{p}$ and $\ket{d - 1}$ for $p$ satisfying $0\leq p <d-1$.
	
In the first step of the protocol, the input photons are pairwise combined on two high-dimensional (HD) beam splitters such that photons~$1$ and~$2$, as well as photons~$3$ and~$4$ interfere with each other, respectively. 
Each of the HD beam splitters has two input ports, labelled~A and~B, as well as two output ports labelled~C and~D, respectively. We choose the labels such that the input port~A receives the input photons~$1$ and~$4$, while the input port~B receives the auxiliary photons~$2$ and~$3$.
	
The HD beam splitters are configured such that photons in the state $\ket{d - 1}$ enter from input port~A (B), and then exit from port~D (C); while photons in the other states $\ket{0}, \ \ldots, \ket{d-2}$ entering input port~A (B), exit from port~C (D)~\cite{WangEtAl2018}.
	
Although such HD beam splitters conserve the total photon number (when ignoring losses), the number of photons exiting from any given output port may be zero, one, or two in principle, depending on the input states. However, here we post-select on those cases where one and only one photon exits at each output port. 
With this in mind, we designate the photons exiting from output port~C of the two HD beam splitters as photons~$1$ and~$4$, while the photons exiting from output port~D are designated as photons~$2$ and~$3$, respectively, as illustrated in Fig.~1 in the main text.

\begin{widetext}	
Including the post-selection, thus the role of the HD beam splitters is to transform the initial state 

\begin{align}
	\left| \psi \right\rangle_{\mathrm{in}} &= \ket{\psi}_{14} \otimes {\ket{\phi}_2} \otimes {\ket{\phi}_3} = \frac{1}{2}
	\sum_{m,n=0}^{d-1}c_{m,n}\ket{m}_{1} \bigl( \ket{p}_{2}+\ket{d - 1}_{2}\bigr)\bigl( \ket{p}_{3}+\ket{d - 1}_{3}\bigr)\ket{n}_{4}\nonumber \\
	&= \frac{1}{2}
	\sum_{m,n =0}^{d-1}c_{m,n}
	\ket{m}_{1} \ket{p}_{2}
	\ket{p}_{3}\ket{n }_{4}
	+ 
	\frac{1}{2}\sum_{m,n =0}^{d-1}c_{m,n }
	\ket{m}_{1} \ket{p}_{2}
	\ket{d - 1}_{3}\ket{n }_{4}
	\nonumber\\
	&\ \ \  + 
	\frac{1}{2}\sum_{m,n =0}^{d-1}c_{m,n }
	\ket{m}_{1} \ket{d - 1}_{2}
	\ket{p}_{3}\ket{n }_{4} + 
	\frac{1}{2}\sum_{m,n =0}^{d-1}c_{m,n }
	\ket{m}_{1} \ket{d - 1}_{2}
	\ket{d - 1}_{3}\ket{n }_{4}
\end{align}
into the output state
\begin{align}
	\left| \psi \right\rangle_{\mathrm{out}} &= \sum_{m,n =0}^{d-2}c_{m,n }\ket{m}_{1} \ket{p}_{2}\ket{p}_{3}\ket{n }_{4}
	+\!\sum_{k=0}^{d-2}c_{m,d-1}\ket{m}_{1} 
	\ket{p}_{2}\ket{d - 1}_{3}\ket{d - 1}_{4}  
	\nonumber\\
	& \ \ \ +\!\sum_{n =0}^{d-2}c_{d-1,n }\ket{d - 1}_{1} 
	\ket{d - 1}_{2}\ket{p}_{3}\ket{n }_{4} + c_{d-1,d-1}\ket{d - 1}_{1} \ket{d - 1}_{2}\ket{d - 1}_{3}\ket{d - 1}_{4} ,
\end{align}  
which is normalized because 
\begin{align}
	\sum_{m,n =0}^{d-2}|c_{m,n }|^{2}
	+ 
	\sum_{m=0}^{d-2}|c_{m,d-1}|^{2}
	+ 
	\sum_{n =0}^{d-2}|c_{d-1,n }|^{2}
	+ |c_{d-1,d-1}|^{2}
	= 
	\sum_{m,n =0}^{d-1}|c_{m,n }|^{2} = 1 .
\end{align}

Finally, we apply a Hadamard gate in the two-dimensional subspace spanned by the states $\ket{p}_{3}$ and $\ket{d - 1}_{3}$ of photon~$3$ exiting from output port D of the second HD beam splitter, explicitly, the operation acts on the Hilbert space of photon~$3$ as
\begin{align}
	H_{3}  &= \frac{1}{\sqrt{2}} \Bigl( 
	\ket{p}\!\bra{p}_{3} + 
	\ket{p}\!\bra{d-1}_{3} + 
	\ket{d - 1}\!\bra{p}_{3} - 
	\ket{d - 1}\!\bra{d-1}_{3}
	\Bigr) ,
\end{align}
resulting in the four-photon state
\begin{align}
	H_{3} \left| \psi \right\rangle_{\mathrm{out}} &= \frac{1}{\sqrt{2}} \sum_{m,n =0}^{d-2}c_{m,n }\ket{m}_{1} \ket{p}_{2}\bigl( \ket{p}_{3}+\ket{d - 1}_{3}\bigr)\ket{n }_{4}
	+\frac{1}{\sqrt{2}}\sum_{m=0}^{d-2}c_{m,d-1}\ket{m}_{1} 
	\ket{p}_{2} \bigl( \ket{p}_{3}-\ket{d - 1}_{3}\bigr) \ket{d - 1}_{4}\nonumber\\
	& 
	\ \ \ +\frac{1}{\sqrt{2}}\sum_{n =0}^{d-2}c_{d-1,n }\ket{d - 1}_{1} 
	\ket{d - 1}_{2}\bigl( \ket{p}_{3}+\ket{d - 1}_{3}\bigr)\ket{n }_{4}
	+\frac{c_{d-1,d-1}}{\sqrt{2}}\ket{d - 1}_{1} \ket{d - 1}_{2} \bigl( \ket{p}_{3}-\ket{d - 1}_{3}\bigr) \ket{d - 1}_{4}.
\end{align}
We can then rewrite the above state using the Bell-state basis consisting of the states $ \ket{\Phi^{\pm}}_{23} = \tfrac{1}{\sqrt{2}} \bigl( \ket{p}_{2}\ket{p}_{3} \pm \ket{d - 1}_{2}\ket{d - 1}_{3}\bigr)$ and $\ket{\Psi^{\pm}}_{23} = \frac{1}{\sqrt{2}} \bigl( \ket{p}_{2}\ket{d - 1}_{3} \pm \ket{d - 1}_{2}\ket{p}_{3}\bigr)$ in the four-dimensional subspace spanned by the states $\ket{m}_{2}\ket{n}_{3}$ for $(m,n=p,d - 1)$ of photons~$2$ and~$3$, that is,
\begin{align}
	H_{3} \left| \varphi \right\rangle_{\mathrm{out}} & = \frac{1}{\sqrt{2}} \sum_{m,n =0}^{d-2}c_{m,n }\ket{m}_{1} 
	\Bigl( \ket{\Phi^{+}}_{23}+\ket{\Phi^{-}}_{23}
	+\ket{\Psi^{+}}_{23}+\ket{\Psi^{-}}_{23}\Bigr)
	\ket{n }_{4}\nonumber\\
	& \quad  +\frac{1}{\sqrt{2}}\sum_{m=0}^{d-2}c_{m,d-1}\ket{m}_{1} 
	\Bigl( \ket{\Phi^{+}}_{23}+\ket{\Phi^{-}}_{23}
	-\ket{\Psi^{+}}_{23}-\ket{\Psi^{-}}_{23}\Bigr)
	\ket{d - 1}_{4}\nonumber\\
	& \quad +\frac{1}{\sqrt{2}}\sum_{n =0}^{d-2}c_{d-1,n }\ket{d - 1}_{1} 
	\Bigl( \ket{\Phi^{+}}_{23}-\ket{\Phi^{-}}_{23}
	+\ket{\Psi^{+}}_{23}-\ket{\Psi^{-}}_{23}\Bigr)
	\ket{n }_{4}\nonumber\\
	& \quad +\frac{1}{\sqrt{2}}c_{d-1,d-1}\ket{d - 1}_{1} 
	\Bigl(-\ket{\Phi^{+}}_{23}+\ket{\Phi^{-}}_{23}
	+\ket{\Psi^{+}}_{23}-\ket{\Psi^{-}}_{23}\Bigr)
	\ket{d - 1}_{4}\nonumber\\
	& = \frac{1}{2}\Bigl( 
	\sum_{m,n =0}^{d-2}c_{m,n }\ket{m}_{1}\ket{n }_{4}
	+ 
	\sum_{m=0}^{d-2}c_{m,d-1}\ket{m}_{1}\ket{d - 1}_{4}
	+ 
	\sum_{n =0}^{d-2}c_{d-1,n }\ket{d - 1}_{1}\ket{n }_{4}
	- c_{d-1,d-1}\ket{d - 1}_{1}\ket{d - 1}_{4}\Bigl) \ket{\Phi^{+}}_{23}\nonumber\\
	&+ \frac{1}{2}\Bigl( 
	\sum_{m,n =0}^{d-2}c_{m,n }\ket{m}_{1}\ket{n }_{4}
	+ 
	\sum_{m=0}^{d-2}c_{m,d-1}\ket{m}_{1}\ket{d - 1}_{4}
	- 
	\sum_{n =0}^{d-2}c_{d-1,n }\ket{d - 1}_{1}\ket{n }_{4}
	+ c_{d-1,d-1}\ket{d - 1}_{1}\ket{d - 1}_{4}\Bigl) \ket{\Phi^{-}}_{23}\nonumber\\
	&+ \frac{1}{2}\Bigl( 
	\sum_{m,n =0}^{d-2}c_{m,n }\ket{m}_{1}\ket{n }_{4}
	- 
	\sum_{m=0}^{d-2}c_{m,d-1}\ket{m}_{1}\ket{d - 1}_{4}
	+ 
	\sum_{n =0}^{d-2}c_{d-1,n }\ket{d - 1}_{1}\ket{n }_{4}
	+ c_{d-1,d-1}\ket{d - 1}_{1}\ket{d - 1}_{4}\Bigl) \ket{\Psi^{+}}_{23}\nonumber\\
	&+ \frac{1}{2}\Bigl( 
	\sum_{m,n =0}^{d-2}c_{m,n }\ket{m}_{1}\ket{n }_{4}
	- 
	\sum_{m=0}^{d-2}c_{m,d-1}\ket{m}_{1}\ket{d - 1}_{4}
	- 
	\sum_{n =0}^{d-2}c_{d-1,n }\ket{d - 1}_{1}\ket{n }_{4}
	- c_{d-1,d-1}\ket{d - 1}_{1}\ket{d - 1}_{4}\Bigl) \ket{\Psi^{-}}_{23}.
\end{align} 
By defining the unitary correction operators $U_{1} = I - 2 \ket{d - 1} \bra{d - 1}_{1}$ and $U_{4}= I - 2 \ket{d - 1} \bra{d - 1}_{4}$, as well as the CPF gate $U_{\mathrm{CPF}} = I - 2 \ket{d - 1}\ket{d - 1}_{1} \otimes \bra{d-1}\bra{d - 1}_{4}$, the final state $\left| \psi \right\rangle_{\mathrm{f}}\equiv H_{3} \left| \psi \right\rangle_{\mathrm{out}}$ can be written more compactly as
\begin{align}
	\left| \psi \right\rangle_{\mathrm{f}} = U_{\mathrm{CPF}}\ket{\psi}_{14} \ket{\Phi^{+}}_{23} + {U_{1}}U_{\mathrm{CPF}}\ket{\psi}_{14} \ket{\Phi^{-}}_{23} + {U_{4}}U_{\mathrm{CPF}}\ket{\psi}_{14} \ket{\Psi^{+}}_{23} + {U_{1}}{U_{4}}U_{\mathrm{CPF}}\ket{\psi}_{14} \ket{\Psi^{-}}_{23}.
\end{align}

Thus we see that a Bell-state measurement in basis $\left\lbrace \ket{\Phi^{\pm}}_{23},\ket{\Psi^{\pm}}_{23}\right\rbrace $ of photons~$2$ and~$3$, and some suitable local corrections (none, $U_{1}$, $U_{4}$, or $U_{1}U_{4}$) depending on the outcome, result in the application of the CPF gate to the two-photon input state $\ket{\psi}_{14}$.	
\end{widetext}
	
\section{Implementation of CPF Gate Using OAM}
	\label{sec:OAM implementation}
	
In this section, we discuss how our protocol for realizing the CPF gate is implemented in dimension $d=4$ by encoding the qubits in the photonic orbital angular momentum (OAM). We begin by describing the preparation of photons representing the initial states of the control and target qudits, as well as those representing the auxiliary qudits in Sec.~\ref{subsec:Preparation of control and target states} and Sec.~\ref{subsec:preparation of the auxiliary states}, respectively. In Sec.~\ref{subsec:OAM HD beam splitter} we describe the implementation of the HD beam splitters using OAM and polarization, while Sec.~\ref{subsec:Hadamard gate and Bell-state measurement} provides the details for the realization of the Hadamard gate and Bell-state measurements.
	
	
\subsection{Preparation of control and target states}\label{subsec:Preparation of control and target states}
	
In our experimental realization for $d=4$, we represent the computational-basis states $\ket{0}$, $\ket{1}$, $\ket{2}$, and $\ket{3}$ by the OAM states $\ketOAM{-2}$, $\ketOAM{-1}$, $\ketOAM{0}$, and $\ketOAM{+1}$, respectively, where $\ketOAM{\ell}$ denotes the photon being in the Laguerre-Gauss mode with radial index zero and azimuthal quantum number~$\ell$. 
	
As described in more detail in Sec.~\ref{sec:bases used to estimate the fidelity of the CPF gate}, the control and target qubits, encoded in photons $1$ and $4$, respectively. Each of the photons $1$ and $4$ can be prepared in any one of ten different states: four OAM states with different azimuthal quantum number~$\ell$, i.e., the four computational-basis states $\ketOAM{-2}$, $\ketOAM{-1}$, $\ketOAM{0}$, and $ \ketOAM{+1}$, four equally weighted superpositions of pairs of OAM states with azimuthal quantum numbers of the same parity, i.e., $\frac{1}{\sqrt{2}}\bigl(\ketOAM{-2}\pm\ketOAM{0}\bigr)$ and $\frac{1}{\sqrt{2}}\bigl(\ketOAM{-1}\pm \ketOAM{+1}\bigr)$, as well as two equally weighted superpositions of computational-basis states, $\frac{1}{\sqrt{2}}\bigl(\ketOAM{-1}+\ketOAM{0}\bigr)$ and $\frac{1}{\sqrt{2}}\bigl(\ketOAM{0}+ \ketOAM{+1}\bigr)$. 
For each of these states, the corresponding photons were first prepared by passing a Gaussian beam ($\ell=0$) through a polarizing beam splitter (PBS), which transmits horizontally polarized photons and reflects vertically polarized ones. The initial state $\ket{H}\ketOAM{0}$ is then subjected to a sequence of optical elements: The half-wave plates (HWPs) rotate the linear polarization by an angle $2\alpha$ in the plane perpendicular to the beam propagation, i.e.,
\begin{subequations}
	\label{eq:HWP}
	\begin{align}
	\operatorname{HWP}(\alpha) \ket{H} &= 
	\cos(2\alpha) \ket{H} + \sin(2\alpha)\ket{V}, \\
	\operatorname{HWP}(\alpha) \ket{V} &= 
	\sin(2\alpha) \ket{H} - \cos(2\alpha)\ket{V}.
	\end{align}
\end{subequations}
The quarter-wave plates (QWPs), whose action for relevant selected angles is described by
\begin{subequations}
	\label{eq:QWP pi over 4}
	\begin{align}
	\operatorname{QWP}(\pi/4) \ket{H} &= 
	e^{i\pi/4} \frac{1}{\sqrt{2}} \bigl( \ket{H} + i \ket{V}\bigr) = e^{i\pi/4} \ket{R}, \\
	\operatorname{QWP}(\pi/4) \ket{V} &= 
	- e^{-i\pi/4} \frac{1}{\sqrt{2}} \bigl( \ket{H} - i \ket{V}\bigr) = - e^{-i\pi/4} \ket{L}, \\
	\operatorname{QWP}(-\pi/4) \ket{H} &= 
	e^{i\pi/4} \frac{1}{\sqrt{2}} \bigl( \ket{H} - i \ket{V}\bigr) = e^{i\pi/4} \ket{L}, \\
	\operatorname{QWP}(-\pi/4) \ket{V} &= 
	e^{i\pi/4} \frac{1}{\sqrt{2}} \bigl( \ket{H} + i \ket{V}\bigr) = e^{i\pi/4} \ket{R},
	\end{align}
\end{subequations}
switch between linear polarization, $\ket{H}$ and $\ket{V}$, and right- and left-circular polarization, $\ket{R}=\tfrac{1}{\sqrt{2}} \bigl( \ket{H} + i \ket{V}\bigr)$ and $\ket{L}=\tfrac{1}{\sqrt{2}} \bigl( \ket{H} - i \ket{V}\bigr)$, respectively. 
The q-plates (QPs) act jointly on the polarization and OAM according to~\cite{MarrucciManzoPaparo2006} 
\begin{subequations}
	\begin{align}
	\operatorname{QP}(q) \ket{R}\ketOAM{\ell} &= \ket{L}\ketOAM{\ell+2q} ,\\
	\operatorname{QP}(q) \ket{L}\ketOAM{\ell} &= \ket{R}\ketOAM{\ell-2q} ,
	\end{align}
\end{subequations}
and here we use two different types of such devices: QPs with $q=1/2$ create the superpositions of states with azimuthal quantum numbers differing by $\Delta\ell=1$, 
while QPs with $q=1/4$ allow us to create superpositions of states with azimuthal quantum numbers differing by $\Delta\ell=1/2$.
States with such fractional azimuthal quantum numbers are coherent superpositions of an infinite number of OAM state with integer azimuthal quantum numbers (which form a basis of the infinite-dimensional Hilbert space describing a photon in the paraxial approximation)~\cite{PhysRevLett.95.240501}, and can be either generated by spatial light modulators~\cite{Wang:07, zhou2016experimental} or a simple elements such as a QP with $q=1/4$ to obtain $\Delta \ell =2q= 1/2$. 
A spiral phase plate (SPP) can reduce the azimuthal quantum number by one, $\operatorname{SPP}\ketOAM{\ell}=\ketOAM{\ell-1}$, while we use PBSs to post-select on the output corresponding to horizontal polarization when required. The sequences of these operations generating the desired (superpositions of) OAM states are shown in Table~\ref{TableS1}.	
\begin{table*}[]
	\centering
	\caption{Preparation of OAM states with a single value of $\ell$ and superpositions of OAM states with $\Delta\ell=2$. In our setup, a PBS is used to create the initial state $\ket{H}\ketOAM{0}$, followed by a sequence of HWPs, QWPs, a QP, and a spiral phase plate (SPP), as well as another PBS (and post-selection on $\ket{H}$ at its output port), to prepare the OAM states with a single value of $\ell$ and superpositions of OAM states with $\Delta\ell=2$ listed in the last column on the right-hand side. Each row, from left to right, lists the parameters for the respective elements, with ``---'' indicating the absence of the corresponding element.}
    \label{TableS1}
	\begin{tabular}{ccccccccc}
	\hline
	\hline
	\ initial state\ \    &\ HWP($\alpha_1$)\ \ &\ QWP($\beta_1$)\ \ &\ QP($q$)\ \ &\ QWP($\beta_2$)\ \ &\ HWP($\alpha_2$)\ \ &\ SPP($\Delta\ell$)\ \ &\ PBS\ \ &\ prepared OAM state\ \\
	\hline
	\ $\ket{H}\ketOAM{0}$ &\ ---\ \ & $-\pi/4$ & $1/2$ & -$\pi/4$ &\ ---\ \ & $-1$ & \ $\ket{H}\!\bra{H}$\ \ & $\ketOAM{-2}$\\ 
	\ $\ket{H}\ketOAM{0}$ &\ ---\ \ & $-\pi/4$ & $1/2$ & -$\pi/4$ &\ ---\ \ & --- & \ $\ket{H}\!\bra{H}$\ \ & $\ketOAM{-1}$\\ 
	\ $\ket{H}\ketOAM{0}$ &\ ---\ \ & $\phantom{-}\pi/4$ & $1/2$ & $\phantom{-}\pi/4$ &\ ---\ \ & $-1$ & \ $\ket{H}\!\bra{H}$\ \ & $\ketOAM{0}$\\ 
	\ $\ket{H}\ketOAM{0}$ &\ ---\ \ & $\phantom{-}\pi/4$ & $1/2$ & $\phantom{-}\pi/4$ &\ ---\ \ & --- & \ $\ket{H}\!\bra{H}$\ \ & $ \ketOAM{+1}$\\
	\ $\ket{H}\ketOAM{0}$ & $\phantom{-}\pi/8$ & $\phantom{-}\pi/4$ & $1/2$ & $\phantom{-}\pi/4$ & $\pi/8$ & $-1$ &\ $\ket{H}\!\bra{H}$\ \ & \ \ $\frac{1}{\sqrt{2}} \bigl( \ketOAM{-2}+\ketOAM{0}\bigr)$\ \ \ 
	\\
	\ $\ket{H}\ketOAM{0}$ & $\phantom{-}\pi/8$ & $\phantom{-}\pi/4$ & $1/2$ & $\phantom{-}\pi/4$ & $\pi/8$ &\ ---\ \ &\ $\ket{H}\!\bra{H}$\ \ & \ \ $\frac{1}{\sqrt{2}} \bigl( \ketOAM{-1}+ \ketOAM{+1}\bigr)$\ \ \ 
	\\
	\ $\ket{H}\ketOAM{0}$ & $-\pi/8$ & $\phantom{-}\pi/4$ & $1/2$ & $\phantom{-}\pi/4$ & $\pi/8$ & $-1$ &\ $\ket{H}\!\bra{H}$\ \ & \ \ $\frac{1}{\sqrt{2}} \bigl( \ketOAM{-2}-\ketOAM{0}\bigr)$\ \ \ 
	\\
	\ $\ket{H}\ketOAM{0}$ & $-\pi/8$ & $\phantom{-}\pi/4$ & $1/2$ & $\phantom{-}\pi/4$ & $\pi/8$ &\ ---\ \ &\ $\ket{H}\!\bra{H}$\ \ & \ \ $\frac{1}{\sqrt{2}} \bigl( \ketOAM{-1}- \ketOAM{+1}\bigr)$\ \ \ 
	\\
	\hline
	\hline
    \ initial state\ \    &\ QP($q_1$)\ \ &\ PBS\ \ &\ QWP($\beta_1$)\ \ &\ QP($q_2$)\ \ &\ QWP($\beta_2$)\ \ &\ SPP($\Delta\ell$)\ \ &\ PBS\ \ &\ prepared OAM state\ \\ \hline
	\ $\ket{H}\ketOAM{0}$ &\ 1/4 \ \ & $\ket{H}\!\bra{H}$ & $\pi/4$ & 1/4 &\ $\pi/4$ \ \ & --- & \ $\ket{H}\!\bra{H}$\ \ & $\frac{1}{\sqrt{2}} \bigl(\ketOAM{0}+ \ketOAM{+1}\bigr)$\\ 
	\ $\ket{H}\ketOAM{0}$ &\ 1/4 \ \ & $\ket{H}\!\bra{H}$ & $\pi/4$ & 1/4 &\ $\pi/4$ \ \ & $-1$ & \ $\ket{H}\!\bra{H}$\ \ & $\frac{1}{\sqrt{2}} \bigl(\ketOAM{-1}+\ketOAM{0}\bigr)$
	\\
	\hline
	\hline	
	\end{tabular}
\end{table*}	
	
\subsection{Preparation of the auxiliary states}\label{subsec:preparation of the auxiliary states}
	
The two auxiliary photons~$2$ and~$3$ output from a single-mode fiber (SMF) are also initially prepared in the state $\ket{H}\ketOAM{0}$. 
In the general scheme described in Fig.~2 of the main text, these auxiliary photons should enter the HD beam splitters at the respective ports B in the states $\frac{1}{\sqrt{2}}(\ketOAM{-1}+ \ketOAM{+1})$. There, the photons successively pass through a sequence of optical elements, which should result in the state $\frac{1}{\sqrt{2}}(\ket{V}\ketOAM{-1}+\ket{H} \ketOAM{+1})$ [as we will see in the first line of Eq.~(\ref{eq:input port b})] before being combined with photons from other input ports at PBS2. 
However, in the experimental setup, illustrated in Fig.~3 of the main text, we chose to forgo the preparation of $\frac{1}{\sqrt{2}}(\ketOAM{-1}+ \ketOAM{+1})$ before switching to $\frac{1}{\sqrt{2}}(\ket{V}\ketOAM{-1}+\ket{H} \ketOAM{+1})$. Instead, we prepared the state $\frac{1}{\sqrt{2}}(\ket{V}\ketOAM{-1}+\ket{H} \ketOAM{+1})$ directly using a QP ($q=1/2$) and a QWP ($\pi/4$), such that the initial state $\ket{H}\ketOAM{0}$ is transformed according to
\begin{align}
\ket{H}\ketOAM{0} &	= \frac{1}{{\sqrt 2 }} { \left( \ket{L} + \ket{R} \right) } \ketOAM{0} \nonumber \\
& \xrightarrow{\text{QP($q=1/2$)}}\frac{1}{{\sqrt 2 }} { \left( \ket{R} \ketOAM{-1} + \ket{L} \ketOAM{+1} \right) } \nonumber \\
& \xrightarrow{\text{QWP}(\pi/4)}\frac{1}{{\sqrt 2 }} { \left( \ket{V} \ketOAM{-1} + \ket{H} \ketOAM{+1} \right) }.
\end{align}
	
	
\subsection{OAM HD beam splitter}\label{subsec:OAM HD beam splitter}
	
We now discuss the realization of the OAM HD beam splitter as shown in Fig.~2 of the main text. The setup consists of operations we refer to as CNOTs of order~$k$, or O$_{k}$-CNOTs (with $k=1$ or $2$), whose composition and function we will explain shortly. In addition the HD beam splitters contain PBSs and HWPs (see Sec.~\ref{subsec:Preparation of control and target states}) as well as Dove prisms (DPs), phase plates (PPs), delay lines (DLs), and mirrors. We shall now proceed to explain these elements.
To begin we recall that PBSs are used to separate input photons with horizontal and vertical polarizations into different output paths. 
Specifically, the horizontal polarization is transmitted, while the vertical polarization is reflected and introduces a phase shift of $\pi/2$. 
	
The core components of the OAM HD beam splitter are the O$_{k}$-CNOTs, whose role is to correlate the parity of the azimuthal quantum number $\ell$ with the (linear) polarization of the photons such that different values of $\ell$ can be separated with the aid of PBSs. 
Each O$_{k}$-CNOT consists of an interferometer between two HWPs (or a HWP and a QWP), as shown in Fig.~3 of the main text. The interferometers themselves each include two DPs and a HWP. 
	
For the O$_{1}$-CNOT, in the interferometer the angles of the two DPs  are $\pm\pi/4$ and the HWP is at an angle $\pi/4$. 
Photons passing through the interferometer hence transform as
\begin{subequations}
	\begin{align}
	\ket{H} \ketOAM{\ell} &\to {e^{i\ell\pi/2 }}\ket{V} \ketOAM{-\ell}, \\
	\ket{V} \ketOAM{\ell} &\to {e^{-i\ell\pi/2 }}\ket{H} \ketOAM{-\ell}, 
	\end{align}
\end{subequations}
where the DP changes the sign of the azimuthal quantum number $\ell$ but also imparts a phase shift that depends on $\ell$ and on the DP's angle of $\gamma$ according to
\begin{align}
	\operatorname{DP} \ketOAM{\ell} &= i e^{2 i \gamma \ell} \ketOAM{-\ell} .
\end{align}
In the O$_{1}$-CNOT, the interferometer is placed between two HWPs with angles of $\pi/8$, such that the total transformation of the O$_{1}$-CNOT has been given in Eq.~(\ref{eq:4}) of the main text.
		
For the O$_{2}$-CNOTs, the angles of the two DPs are $\pm\pi/8$ and the HWP is at $\pi/4$ in the interferometer, such that the latter transforms input states according to
\begin{subequations}
	\begin{align}
	\ket{H} \ketOAM{\ell} &\to {e^{i\ell\pi/4 }}\ket{V} \ketOAM{-\ell} ,\\
	\ket{V} \ketOAM{\ell} &\to {e^{-i\ell\pi/4 }}\ket{H} \ketOAM{-\ell} .
	\end{align}
\end{subequations}
In addition, the O$_{2}$-CNOT interferometer is placed between a HWP at an angle of $\pi/8$ in the front and a QWP at an angle of $\pi/4$ in the back. The entire O$_{2}$-CNOT thus transforms the polarization and azimuthal quantum number of photons passing through it according to
\begin{widetext}
\begin{subequations}
	\begin{align}
	\text{O$_{2}$-CNOT} \ket{H} \ketOAM{\ell} & = \phantom{- }
	e^{-i (\ell-1)\pi /4} 
	\left[ 
	\frac{1-e^{i (\ell-1)\pi /2}}{2} \ket{H}
	+ i 
	\frac{1+e^{i (\ell-1)\pi /2}}{2} \ket{V}
	\right] \ketOAM{-\ell} ,\\
	\text{O$_{2}$-CNOT} \ket{V} \ketOAM{\ell} & = 
	- e^{-i (\ell-1)\pi /4} 
	\left[ 
	\frac{1+e^{i (\ell-1)\pi /2}}{2} \ket{H}
	+ i 
	\frac{1-e^{i (\ell-1)\pi /2}}{2} \ket{V}
	\right] \ketOAM{-\ell} .
	\end{align}
\end{subequations}
\end{widetext}
	
To realize the OAM HD beam splitter for dimension of $d=4$, the O$_{k}$-CNOTs are combined with PBSs (splitting the linear polarizations into different paths), mirrors (imparting a $\pi/2$ phase shift), HWPs [rotating the linear polarization, see Eq.~(\ref{eq:HWP})], DPs, PPs, and DLs as shown in Fig.~2 of the main text. 
Photons incident on input port A then undergo the following sequence of transformations, where the different paths taken are represented by states $\ket{\mathrm{A}}$, $\ket{\mathrm{C}}$, $\ket{\mathrm{D}}$, $\ket{\mathrm{P}1}$, and $\ket{\mathrm{P}2}$, respectively, and we thus have
\begin{widetext}
\begin{align}
	& \ket{H} \bigl( a_{-2} \ketOAM{-2}+a_{-1} \ketOAM{-1}+a_{0} \ketOAM{0}+a_{1} \ketOAM{+1} \bigr) \ket{\mathrm{A}} \nonumber\\
	& \hspace*{5mm}\xrightarrow{\mathrm{ O_1-CNOT }}\ 
	\ket{H}\left(e^{i \pi} a_{-2} \ketOAM{2}+a_{0} \ketOAM{0}\right) +\ket{V}\left(e^{i \pi / 2} a_{-1} \ketOAM{+1}+e^{-i \pi / 2} a_{1} \ketOAM{-1}\right) \ket{\mathrm{A}} \nonumber\\
	& \hspace*{5mm}\xrightarrow{\mathrm{ PBS_1 }}\ 
	\ket{H}\left(e^{i \pi} a_{-2} \ketOAM{2}+a_{0} \ketOAM{0}\right) \ket{\mathrm{P}1}
	+ e^{i \pi / 2}\ket{V}\left(e^{i \pi / 2} a_{-1} \ketOAM{-1}+e^{-i \pi / 2} a_{1} \ketOAM{+1}\right) \ket{\mathrm{P}2} \nonumber\\
	& \hspace*{5mm}\xrightarrow{(\mathrm{ O_2-CNOT })_{\mathrm{P}2}}\ \ket{H}\left(e^{i \pi} a_{-2} \ketOAM{2}+a_{0} \ketOAM{0}\right) \ket{\mathrm{P}1}
	+\left(-a_{-1}\ket{V} \ketOAM{+1}+e^{i \pi} a_{1} \ket{H}\ketOAM{-1} \right) \ket{\mathrm{P}2}\nonumber \\
	& \hspace*{5mm}\xrightarrow{\mathrm{PBS2}}\ 
	\ket{H}\left(e^{i \pi} a_{-2} \ketOAM{2}+a_{0} \ketOAM{0}\right) \ket{\mathrm{P}1}
	-e^{i \pi / 2} a_{-1}\ket{V}\ketOAM{-1} \ket{\mathrm{P}2}
	- a_{1}\ket{H}\ketOAM{-1} \ket{\mathrm{D}} \nonumber\\
	& \hspace*{5mm}\xrightarrow{(\mathrm{ O_2-CNOT } \text{\& mirror}\text{ \& PP})_{\mathrm{P}2}
	}\ 
	\ket{H}\left(e^{i \pi} a_{-2} \ketOAM{2}+a_{0} \ketOAM{0}\right) \ket{\mathrm{P}1}
	+e^{i \pi } a_{-1}\ket{V}\ketOAM{-1} \ket{\mathrm{P}2}
	- a_{1}\ket{H}\ketOAM{-1} \ket{\mathrm{D}} \nonumber\\
	& \hspace*{5mm}\xrightarrow{\mathrm{ PBS3 }}\ 
	\Bigl[ \ket{H}\left(e^{i \pi} a_{-2} \ketOAM{2}+a_{0} \ketOAM{0}\right) 
	+e^{-i \pi / 2} a_{-1}\ket{V} \ketOAM{+1} \Bigr]\ket{\mathrm{C}}
	- a_{1}\ket{H}\ketOAM{-1}\ket{\mathrm{D}} \nonumber\\
	& \hspace*{5mm}\xrightarrow[{[\mathrm{mirror} \text{ \& } \mathrm{O_2-CNOT} \text{ \& } \mathrm{HWP}(\pi / 4)\text{ \& }\mathrm{DP2}(\pi/4)]_\mathrm{D}}]{(\mathrm{O_1-CNOT})_\mathrm{C}}\ \ket{H}\bigl(a_{-2} \ketOAM{-2}+a_{0} \ketOAM{0}+a_{-1} \ketOAM{-1}\bigr) \ket{\mathrm{C}}+a_{1}\ket{H} \ketOAM{+1}\ket{\mathrm{D}}.	
\end{align}
	
Similarly, photons incident on input port B undergo the sequence of transformations
\begin{align}
	&\ket{H} \bigl( b_{-1} \ketOAM{-1}+b_{1} \ketOAM{+1}\bigr)\ket{\mathrm{B}} \nonumber \\
	\ & \hspace*{15mm} 
	\xrightarrow{\mathrm{O_2-CNOT}\text{ \& }\mathrm{HWP}(\pi / 4)\text{ \& }\mathrm{DP1}(0)\text{ \& PP}}\ 
	\left(b_{-1} \ket{V}\ketOAM{-1} + b_{1} \ket{H} \ketOAM{+1}\right)\ket{\mathrm{B}} \nonumber\\
	& \hspace*{15mm}
	\xrightarrow{\mathrm { PBS2 }}\ b_{1} \ket{H} \ketOAM{+1} \ket{\mathrm{P2}}+ i b_{-1} \ket{V} \ketOAM{+1} \ket{\mathrm{D}} \nonumber\\
	& \hspace*{15mm} 
	\xrightarrow{(\mathrm{O_2}-\mathrm{CNOT}\text{ \& mirror}\text{ \& PP})_{\mathrm{P2}}}\ 
	b_{1} \ket{V} \ketOAM{+1} \ket{\mathrm{P}2}+i b_{-1} \ket{V} \ketOAM{+1} \ket{\mathrm{D}} 
	\label{eq:input port b} \nonumber \\
	& \hspace*{15mm} 
	\xrightarrow{\mathrm{ PBS3 }}\ 
	e^{i \pi / 2} b_{1} \ket{V}\ketOAM{-1} \ket{\mathrm{C}}+i b_{-1} \ket{V} \ketOAM{+1}\ket{\mathrm{D}} \nonumber\\
	& \hspace*{15mm}
	\xrightarrow[{[\mathrm{mirror} \text{ \& } \mathrm{O_2-CNOT}\text{ \& }\mathrm{HWP}(\pi / 4)\text{ \& }\mathrm{DP2}(\pi/4)]}_{\mathrm{D}}]{(\mathrm{O_1-CNOT})_{\mathrm{C}}}\ 
	b_{1} \ket{H} \ketOAM{+1} \ket{\mathrm{C}}+b_{-1} \ket{H}\ketOAM{-1} \ket{\mathrm{D}}.
\end{align}

We note here that the very first step of this transformation was realized differently in the experiment, as explained in Sec.~\ref{subsec:preparation of the auxiliary states}.
\end{widetext}	
	
\subsection{Hadamard Gate and Bell-State Measurement}\label{subsec:Hadamard gate and Bell-state measurement}
	
Performing a Hadamard gate operation and a Bell state measurement (BSM) on the polarization degree of freedom is easier than on OAM. Therefore, we must set the azimuthal quantum number to zero, i.e. $\ketOAM{0}$. To do this, along with a Hadamard gate, we use a simple set of a QWP, a QP and another QWP instead of an O$_{2}$-CNOT, HWP, DP and other elements.
More specifically, both QWPs are rotated to $-\pi/4$ and a QP with $q=1/2$ is used for photon~$2$, which leads to the state transformations such as from $\ket{H} \ketOAM{+1}$ to $\ket{V}\ketOAM{0}$; for photon~$3$, a set of a QWP ($-\pi/4$), a QP ($q=1/2$), and another QWP ($0$) performs a Hadamard gate on the state, i.e., $\ket{V}\ketOAM{-1}$ is mapped to $i\ket{A}\ketOAM{0}$, where $\ket{A}=\frac{1}{\sqrt{2}}(\ket{H}-\ket{V})$.
Then, a BSM between photons~$2$ and~$3$ will be easily implemented by a PBS.
	
	
\section{Arbitrary Quantum Interference Phase Locking}\label{sec:arbitrary quantum interference phase locking}
	
The key to the successful realization of our four-dimensional CPF gate is stable interference on the OAM HD beam splitters. Multipath interferometers can be stabilized in a variety of ways. One approach is to design self-stabilizing interference loops in which different interference paths overlap in both space and time, as is the case in the toroidal Sagnac interferometer~\cite{Wang-Pan2015}. Another approach is to include the various components of the interferometer into an integrated device to ensure the stabilization of the interferometer by passive means~\cite{WangEtAl2018}. Both of the above approaches are limited by the specific spatial structure of the interferometer and cannot adjust the interference locally and flexibly.
\begin{table*}[]
	\begin{center}
	\caption{ZX basis and the XZ basis for photons 1 and 4.}
	\label{TableS2}
	\begin{tabular}{ccccc}
	\hline
	\hline
	& \ \ & Photon-1 & \ \ & Photon-4\\
	\hline
	ZX base & \ \ &$\ketOAM { - 2} $, $\ketOAM { - 1} $, $\ketOAM {0 } $, $\ketOAM {1 } $ & \ \ & $\frac{1}{\sqrt 2} \left( \ketOAM { - 2}  \pm \ketOAM {0 } \right)$, $\frac{1}{\sqrt 2} \left( {\ketOAM { - 1} } \pm \ketOAM {1 } \right)$\\
	\hline
	XZ base & \ \ &$\frac{1}{\sqrt 2} \left( \ketOAM { - 2} \pm \ketOAM{ 0 } \right)$, $\frac{1}{\sqrt 2} \left( {\ketOAM { - 1} } \pm \ketOAM {1 } \right)$& \ \ &$\ketOAM { - 2} $, $\ketOAM { - 1} $, $\ketOAM {0 } $, $\ketOAM {1 } $\\
	\hline
	\hline
	\end {tabular}
	\end{center}
\end{table*}			

Here, we choose a third scheme that uses active phase locking. This scheme requires a series of optical and electrical components, including signal generators, photodetectors (PD), mixers, low-pass filters, electro-optic modulators (EOMs)~\cite{WuHuangYangLiuChen2019}, piezoelectric ceramics (PZTs), and servo-control systems.
	
In the modulation module for the locking laser, an overall phase, ${e^{i\vartheta \sin \Omega t}}$, is added to $\ket{H}$, where $\vartheta$ is the modulation depth, and $\Omega$ is the modulation frequency. ${e^{i\vartheta \sin \Omega t}}$ is achieved by applying an alternating load voltage as a sine wave with an amplitude of +5 V on the EOM. After further modulation, the locking laser can finally be represented as
\begin{equation}
	{E_{\mathrm{in}}} = \frac{1}{{\sqrt 2 }}{E_{0H}}{e^{i\omega t}}{e^{i\vartheta \sin \Omega t}}+ \frac{1}{{\sqrt 2 }}{E_{0V}}{e^{i {\omega t } }}.
\end{equation}
When the locking laser enters the OAM HD beam splitter, beams of different polarizations are separated on a PBS, and the relative phase $\zeta$ between the different paths is incorporated in $\ket{V}$,
\begin{align}
	E_H & = \frac{1}{{\sqrt 2 }} E_{0H} e^{i\omega t} e^{i\vartheta \sin \Omega t}, \\
	E_V & = \frac{1}{{\sqrt 2 }} E_{0V} e^{i\left( \omega t + \zeta \right) }.
\end{align}
	
The optical fields of two beams passes through a polarizer ($\pi/4$) at the output of the OAM HD beam splitter directly before PD, as depicted in Fig.~1(a) of the main text. The resultant field describing the two beams is ${E_{\mathrm{out}}} = \frac{1}{{\sqrt 2 }}\left( {{E_H} + j{E_V}} \right)$. Then, the detected intensity of the interfering signal at the output is ${I_{\mathrm{out}}} = \frac{1}{2}{E_{\mathrm{out}}}^ * {E_{\mathrm{out}}}$.
	
The interfering signal is input to a photodetector, and its AC signal output is mixed with a modulating signal of the same frequency $\cos \left( {\Omega t + \tau } \right)$, where $\tau$ is the phase difference between the electrical signal output from the photodetector and the mixed signal. After being filtered by a low-pass filter, an expression for the error signal $I'$ is obtained as
\begin{equation}
I' \propto \vartheta \sin \zeta \sin \tau.
\end{equation}
	
The error signal is then fed to the servo-control system, which outputs a feedback signal. The feedback signal is applied to the piezoelectric ceramics (PZT), which compensates for various phase perturbations in the interferometric loop. By adjusting the DC compensation of the servo-control system, the interferometer can be locked to any phase.
	
Temperature changes in the experiment induce variations of the phases of $\ket{H}$ and $\ket{V}$ by the EOM with a slow relative phase change, which makes stable phase-locking difficult. A Mach-Zehnder (MZ) interferometer is added to compensate for temperature effects. Adding an MZ interferometer forms a phase-locked optical modulator that generates a locking laser. The latter stabilizes the phase lock, and the locking time for locked OAM HD beam splitters exceeds three hours, ensuring the stable operation of our CPF gate, as shown in Fig.~3 of the main text.

\begin{figure}[!b]
	\centering
	\includegraphics[width=1.00\linewidth]{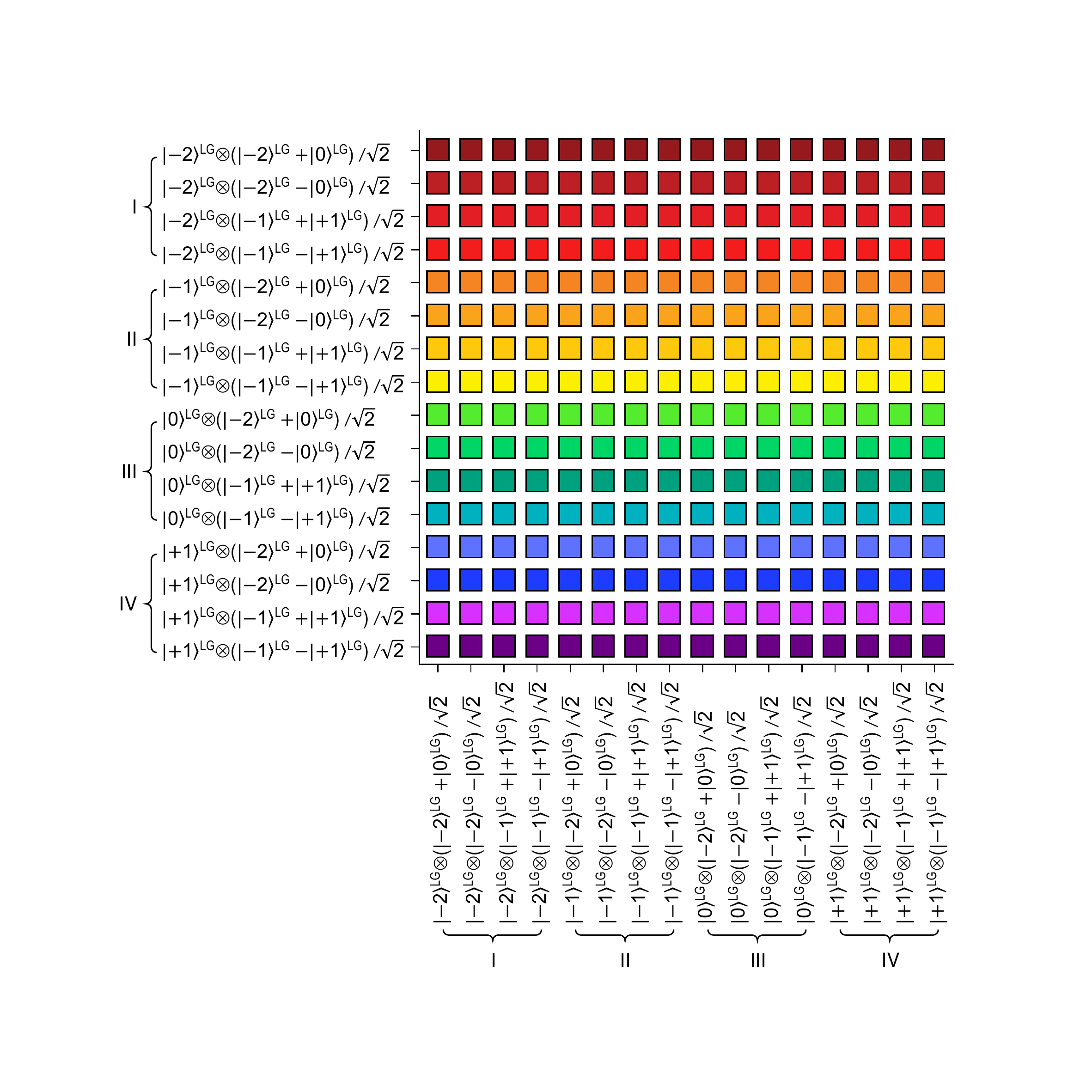}
	\caption{ZX basis for the fidelity measurement of the CPF gate.}
	\label{figS1}
\end{figure}
\begin{figure}[!b]
	\centering
	\includegraphics[width=1.00\linewidth]{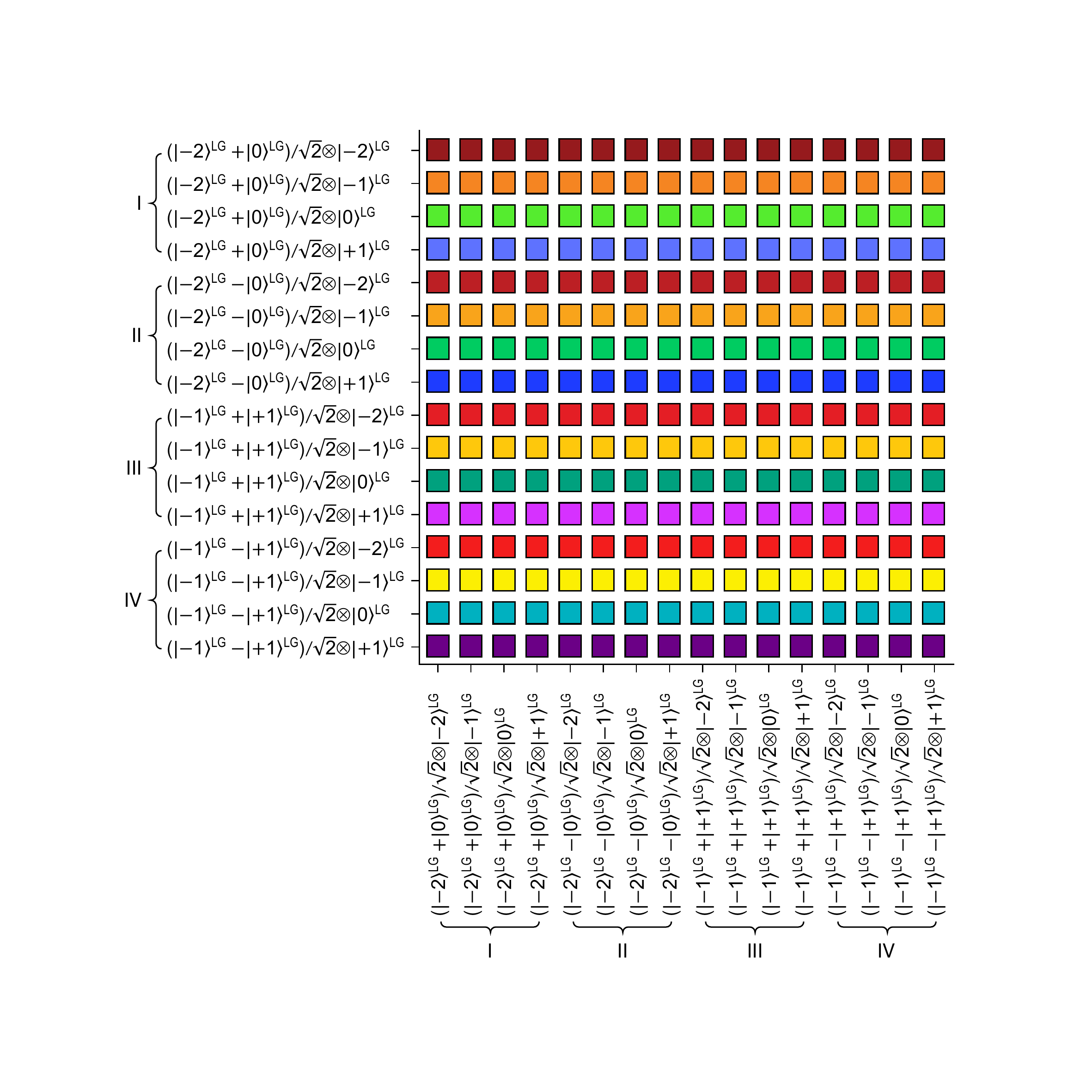}
	\caption{XZ basis for the fidelity measurement of the CPF gate.}
	\label{figS2}
\end{figure}	
	
\section{Bases for Estimation of CPF-Gate Process Fidelity}\label{sec:bases used to estimate the fidelity of the CPF gate}
	
The process fidelity $F$ is a common figure of merit that we use here to benchmark the CPF gate. To calculate the process fidelity $F$ of the CPF gate for two photons encoded in the OAM degree of freedom, we adopt the method proposed in Ref.~\cite{Hofmann2005}, where the process fidelity can be estimated by classical fidelities of two complementary operations. In our work, we chose measurement bases that we refer to here as the ZX basis and the XZ basis of the four-dimensional Hilbert space to estimate the process fidelity $F$ of the experimentally realized CPF gate, as shown in Table~\ref{TableS2}.
		
Each basis contains 16 input states, and we calculate the average probability of the expected output over all of the 16 possible inputs. The corresponding basis in Fig.~4 of the main text is shown in Fig.~\ref{figS1} for the ZX basis and Fig.~\ref{figS2} for the XZ basis).

In addition, we prepared seven superposition states, shown in Table~\ref{TableS3}, to test the CPF gate. 
The details for the preparation of these states are given in Table~\ref{TableS1}.	
\begin{table}[]
	\begin{center}
	\caption{Seven superpositions of OAM states.}
	\label{TableS3}
	\begin{tabular}{ccccc}
	\hline
	\hline
	number & \ \ & photon 1 & \ \ & photon 4\\
	\hline
	1 & \ \ &$\left( {\ketOAM{ { - 2} } + \ketOAM{ 0 } } \right)/\sqrt 2 $ & \ \ &$\left( {\ketOAM{ 0 } + \ketOAM{ 1 } } \right)/\sqrt 2 $\\
	\hline
	2 & \ \ &$\left( {\ketOAM{ { - 2} } + \ketOAM{ 0 } } \right)/\sqrt 2 $& \ \ &$\left( {\ketOAM{ { - 1} } + \ketOAM{ 0 } } \right)/\sqrt 2 $\\
	\hline
	3 & \ \ &$\left( {\ketOAM{ { - 2} } + \ketOAM{ 0 } } \right)/\sqrt 2 $& \ \ &$\left( {\ketOAM{ { - 2} } + \ketOAM{ 0 } } \right)/\sqrt 2 $\\
	\hline
	4 & \ \ &$\left( {\ketOAM{ { - 2} } + \ketOAM{ 0 } } \right)/\sqrt 2 $& \ \ &$\left( {\ketOAM{ { - 1} } + \ketOAM{ 1 } } \right)/\sqrt 2 $\\
	\hline
	5 & \ \ &$\left( {\ketOAM{ { - 1} } + \ketOAM{ 1 } } \right)/\sqrt 2 $& \ \ &$\left( {\ketOAM{ { - 1} } + \ketOAM{ 0 } } \right)/\sqrt 2 $\\
	\hline
	6 & \ \ &$\left( {\ketOAM{ { - 1} } + \ketOAM{ 1 } } \right)/\sqrt 2 $& \ \ &$\left( {\ketOAM{ { - 2} } + \ketOAM{ 0 } } \right)/\sqrt 2 $\\
	\hline
	7 & \ \ &$\left( {\ketOAM{ { - 1} } + \ketOAM{ 1 } } \right)/\sqrt 2 $& \ \ &$\left( {\ketOAM{ { - 1} } + \ketOAM{ 1 } } \right)/\sqrt 2 $\\
	\hline
	\hline
	\end {tabular}
	\end{center}
\end{table}					


\bibliography{Main-SM-0th-Combination}

\end{document}